\documentclass[fleqn,usenatbib]{mnras}

\usepackage{newtxtext,newtxmath}

\usepackage[T1]{fontenc}

\DeclareRobustCommand{\VAN}[3]{#2}
\let\VANthebibliography\thebibliography
\def\thebibliography{\DeclareRobustCommand{\VAN}[3]{##3}\VANthebibliography}

\usepackage{graphicx}	
\usepackage{amsmath}	
\usepackage{xcolor}
\usepackage{bm,url}

\title[Polar dusty circumbinary discs]{Polar alignment of a dusty circumbinary disc -- I. Dust ring formation }

\author[Smallwood et al.]{Jeremy L. Smallwood,$^{1}$\thanks{E-mail: jlsmallwood@asiaa.sinica.edu.tw}
Min-Kai Lin,$^{1,2}$
Hossam Aly,$^{3,4}$
Rebecca Nealon$^{5,6}$
\newauthor
and
Cristiano Longarini$^{7}$
\\
$^{1}$Institute of Astronomy and Astrophysics, Academia Sinica, Taipei 10617, R.O.C.\\
$^{2}${Physics Division, National Center for Theoretical Sciences, Taipei 10617, Taiwan}\\
$^{3}$Faculty of Aerospace Engineering, Delft University of Technology, Kluyverweg 1, 2629 HS Delft, The Netherlands\\
$^{4}$Zentrum für Astronomie der Universität Heidelberg, Astronomisches Rechen-Institut, Mönchhofstr. 12-14, 69120 Heidelberg, Germany \\
$^{5}$Centre for Exoplanets and Habitability, University of Warwick, Coventry CV4 7AL, UK\\
$^{6}$Department of Physics, University of Warwick, Coventry CV4 7AL, UK\\
$^{7}$Dipartimento di Fisica, Università degli Studi di Milano, via Celoria 16, 20133 Milano, Italy\\
}
\date{Accepted XXX. Received YYY; in original form ZZZ}

\pubyear{2024}

\begin{document}
\label{firstpage}
\pagerange{\pageref{firstpage}--\pageref{lastpage}}
\maketitle

\begin{abstract}
We investigate the formation of dust traffic jams in polar-aligning circumbinary discs. We use 3D smoothed particle hydrodynamical simulations of both gas and dust to model an initially highly misaligned circumbinary disc around an eccentric binary. As the circumbinary disc evolves to a polar configuration (perpendicular to the binary orbital plane), the difference in the precession between the gas and dust produces dust traffic jams, which become dense dust rings. We find the formation of dust rings exists for different Stokes number, binary eccentricity, and initial disc tilt. Dust rings are only produced while the circumbinary disc is misaligned to the binary orbital plane. When the disc becomes polar aligned, the dust rings are still present and long-lived. Once these dust rings are formed, they drift inward. The drift timescale depends on the Stokes number. The lower the Stokes number, the faster the dust ring drifts near the inner edge of the disc.  The dust rings will have an increased midplane dust-to-go ratio, which may be a favourable environment for the steaming instability to operate.
\end{abstract}




\section{Introduction}
 In the Galaxy it is estimated that more than $40$-$50\%$ of stars are part of a binary system \citep{Duquennoy1991,Raghavan2010,Tokovinin2014a,Tokovinin2014b}. Young binary star systems are commonly surrounded by a circumbinary disc of gas and dust, which are the sites for planet formation. Observations naturally reveal that circumbinary discs are commonly misaligned to the binary orbital plane \cite[e.g.,][]{Czekala2019}. An initially misaligned circumbinary disc around a circular orbit binary aligns to the binary orbital plane \cite[e.g.,][]{papaloizou1995,Lubow2000,Nixon2011,Facchini2013,Foucart2014}. If the binary eccentricity is nonzero, a low-mass circumbinary disc with a sufficiently large misalignment will precess about the binary eccentricity vector. The angular momentum vector of the disc will align to the binary eccentricity vector, resulting in a polar-aligned circumbinary disc \citep{Aly2015,Martinlubow2017,Martin2018,Lubow2018,Zanazzi2018}. A massive disc aligns to a generalized polar state at lower misalignment to the binary orbital plane \citep{Zanazzi2018,MartinLubow2019}. It is not fully understood how this misalignment affects the gas and dust dynamics and the formation of circumbinary planets.  A binary star system exerts a torque onto the disc, which can impact the planet formation process compared to discs around single stars \citep{Nelson2000,Mayer2005,Boss2006,Martin2014,Fu2015a,Fu2015b,Fu2017}. By understanding the structure and evolution of dusty circumbinary discs, we can shed light on the characteristics of exoplanets.

 The misalignment between a circumbinary disc and binary orbital plane can arise from several mechanisms, including turbulence in star-forming gas clouds \citep{Offner2010,Bate2012,Tokuda2014},  misaligned accretion flows onto young binaries \citep{Bate2010,Bate2018}, and if the angular momentum vector of a binary star is misaligned with
respect to the cloud rotation axis \citep{Bonnell1992}. Binary-disc misalignment occurs in various stages of stellar evolution.  The pre-main-sequence binary KH 15D has a circumbinary disc tilted by about $5^\circ-16^\circ$  \citep{Chiang2004,Smallwood2019,Poon2021}. The circumbinary disc around the young binary IRS 43 has an observed misalignment of about $60^\circ$ \citep{Brinch2016}. There are currently three sources with a polar-aligned, or perpendicular, circumbinary disc. One is the circumbinary disc around HD 98800BaBb \citep{Kennedy2019}, second is around V773 Tau B \citep{Kenworthy2022}, and the third is the $6$--$10\, \rm Gyr$ old binary system, 99 Herculis (99 Her), which has a nearly polar (about $87^\circ$) debris ring \citep{Kennedy2012,Smallwood2020}. Also, the pre-main sequence circumtriple disc around the hierarchical triple star system, GW Ori, is misaligned by about $38^\circ$ \citep{Bi2020,Kraus2020,Smallwood2021c}.

The dynamics of gas and dust in circumbinary discs have yet to be fully understood. In protoplanetary discs, the dust and gas are aerodynamically coupled by drag forces caused by their relative velocities  \citep{Whipple1972,Adachi1976,Weidenschilling1977}. The gas naturally orbits at sub-Keplerian azimuthal velocity because of its (usually negative) pressure gradient and radial velocity due to viscous effects. Contrariwise, dust particles orbit at the Keplerian velocity as it does not experience pressure or viscous forces to a first approximation. Therefore, the dust encounters a headwind within the gaseous disc, causing them to drift inward radially. Depending on the solid particles' size, two different drag force regimes exist. For particles with a size of less than $\sim 1\, \rm m$, the gas mean free path is larger than the dust size ($\lambda_{\rm g} < s$), and the drag force is in the \cite{Epstein1924} regime \cite[e.g.,][]{Laibe2012a}.  In this regime, the fluid is modelled as a collisionless assemblage of molecules with a Maxwellian velocity distribution, and the drag force is proportional to the difference in velocity \citep{Armitage2018}. Nonetheless, dust and gas that are apsidally misaligned may retain a supersonic velocity difference that forces the gas drag to have a  quadratic velocity difference \citep{Kwok1975}. A similar effect may transpire in misaligned discs that undergo nodal precession. A valuable quantity to describe how coupled dust is to the gas in the Epstein regime is by considering the Stokes number given by
\begin{equation}
    \rm St = \frac{\pi}{2} \frac{\rho_{\rm d}s}{\Sigma_{\rm g}},
    \label{eq::st}
\end{equation}
where $\rho_{\rm d}$ is the dust intrinsic density and $\Sigma_{\rm g}$ is the gas surface density. Solid
particles with $\rm St \ll 1$ are strongly coupled to the gas, while the ones with $\rm St \gg 1$ are weakly coupled. The strongest radial drift occurs when $\rm St = 1$. However, particles can become trapped within a local pressure maximum in the gas disc \citep{Nakagawa1986}. Particles' radial drift velocity is zero where the pressure gradient is zero \citep{Rice2006,Zhu2012}.

 Studying the dust dynamics within misaligned circumbinary discs is essential to understanding the general planet formation scenario. \cite{Aly2021} investigated the dynamics of gas and dust in misaligned circumbinary discs using smoothed particle hydrodynamics (SPH). They found that dust pile-ups occur within the dust disc, not due to pressure maxima but to the difference in the precession between the gas and dust \cite[e.g.,][]{Aly2020,Longarini2021}. \cite{Aly2021} simulated circumbinary discs that underwent either coplanar or polar alignment. Their circumbinary disc broke due to the difference in the radial communication and global disc precession timescales for their polar alignment simulations. Therefore, they could not investigate the evolution of the gas and dust in a disc that underwent complete polar alignment.  Dust rings have also been found in simulations of an inclined circumstellar disc undergoing Kozai-Lidov oscillations \citep{Martin2022}. However, these dust rings may be produced primarily by disc eccentricity rather than differential precession.

 The gas disc lifetime around single stars and binary systems can be different. The gas disc lifetimes around single stars are observed to be around $1$-$10\, \rm Myr$ \citep{Haisch2001,Hernandez2007,Hernandez2008,Mamajek2009,Ribas2015}. However, accretion of circumbinary material may be inhibited due to the binary tidal toques, resulting in extended gas disc lifetimes \cite[e.g.,][]{Alexander2012}. For example, the circumbinary gas discs around HD 98800 B, V4046 Sgr, and AK Sco have disc ages of $10 \pm 3$, $23 \pm 3$, and $18 \pm 1\, \rm Myr$, respectively \citep{Soderblom1998,Mamajek2014,Czekala2015}. After the gaseous circumbinary disc is dispersed, the remnant planetesimals produce a second generation of dust through collisions. This leads to the formation of a gas-poor, less massive disc called a debris disc. These debris discs are much cooler in temperature \cite[e.g.,][]{Wyatt2008,Hughes2018}. An initially misaligned disc of debris is unstable. For example, the debris can be modelled as a set of particles on nearly circular ballistic circumbinary orbits that only interact during close encounters or collisions. A particle disc that is initially misaligned with respect to the binary orbital plane by some arbitrary angle will undergo differential nodal precession, leading to spherical distribution of particles and violent collisions \cite[e.g.,][]{Nesvold2016}.  A low-mass exactly polar (or coplanar) debris disc is an exception to this process because it does not undergo nodal precession. 

 A binary system of particular interest is 99 Her, which is encircled by a polar debris disc. Circumbinary dust particles encounter various degrees of coupling to the gas depending on their Stokes number \citep{Birnstiel2010}. Over time, micron-sized dust grains will agglomerate to a higher Stokes number and gradually decouple from the gas disc. If significant decoupling ensues in a misaligned circumbinary disc, the dust particle orbits may evolve independently of the gas disc \citep{Aly2020}. Therefore, to have a stable polar circumbinary debris disc around 99 Her, the primordial gas disc must evolve to a polar configuration before the dust sufficiently decouples. \cite{Smallwood2020} investigated the origin of the polar debris disc around 99 Her. They simulate highly misaligned gas-only circumbinary discs around the 99 Her binary, finding that the polar alignment timescale is much less than the gas disc lifetime. 

 In this work, we further the work of \cite{Aly2021} by investigating polar-aligning coherent dusty circumbinary discs. We model highly misaligned circumbinary discs of gas and dust that undergo alignment to a polar configuration. Unlike the polar simulations by \cite{Aly2021}, we model a rigidly precessing disc that is not susceptible to disc breaking.  The main reason that the polar-aligning discs in \cite{Aly2021} broke compared to the simulations in this work is attributable to the radial extent of the discs. In \cite{Aly2021}, the simulated discs extended outward by a factor of 20 relative to the binary separation, whereas the simulations presented in this work feature discs with a radial extent of 7 times the binary separation. Therefore, we can track the evolution of the gas and dust during the full polar alignment process  without disc breaking.  We find that dust rings are formed in the misaligned disc due to the differences in the gas and dust precession \citep{Aly2020,Longarini2021,Aly2021}. The dust rings evolve to polar with the gas. When the disc is polar, the engine for producing these dust traffic jams is turned off. However, the dust rings produced during the alignment process persist while the disc is polar.  The results from this work have clear implications for the formation of polar circumbinary planets.

 The structure of the paper is as follows. Section~\ref{sec::methods} describes the setup for our hydrodynamical simulations of a misaligned gaseous and dusty circumbinary.  In Section~\ref{sec::hydro_results}, we report the results of our hydrodynamical simulations. In Section~\ref{sec::1D_cal}, we showcase a 1D model used to approximate dust ring formation in a polar-aligning circumbinary disc. We present a discussion in Section~\ref{sec::discussion}.  Lastly, we give a summary in Section~\ref{sec:conclusion}.

\section{Methods}
\label{sec::methods}

We use the 3-dimensional smoothed particle hydrodynamics code {\sc phantom} \citep{Price2018} to model an inclined dusty circumbinary disc. Currently, {\sc phantom}  models the dust-gas mixtures using two techniques depending on the Stokes number of the dust grains. For $\rm St > 1$, two separate types of particles (two-fluid) are simulated, as presented in \cite{Laibe2012a,Laibe2012b}. For smaller grains, $\rm St < 1$, {\sc phantom} uses a single type of particle that represents the combination of dust and gas together (one-fluid), as illustrated in \cite{Price2015a}. With the two-fluid implementation, dust and gas are treated separately with a drag term and explicit timestepping, while with the one-fluid version, dust is considered part of the mixture, with an evolution equations corresponding to the dust fraction. We consider the polar alignment of a circumbinary disc with low gas density and $\rm St > 1$, and therefore use the two-fluid algorithm. The two-fluid implementation takes drag heating into account but neglects thermal coupling between the gas and dust \cite[see][]{Laibe2012a}.

\subsection{Circumbinary disc setup}
We set up a disc of gas and dust particles around an equal-mass binary star system, with $M_1 = M_2 = 0.5\, \rm M_{\odot}$, where $M_1$ is the mass of the primary star and $M_2$ is the mass of the secondary star.  We utilize a Cartesian coordinate system ($x$,$y$,$z$), wherein the $x$-axis is along the direction of the binary eccentricity vector, and the $z$-axis along  with the direction of the binary angular momentum vector.  The disc mass is sufficiently negligible that the binary does not precess, thereby rendering the coordinate system nearly fixed in the inertial frame.  The binary is modelled as sink particles with an initial separation $a_{\rm b} = 16.5\, \rm au$, and an accretion radius of $1.2a_{\rm b} = 20\, \rm au$.  The accretion radius is considered a hard boundary, where particles that penetrate this boundary deposit their mass and angular momentum onto the sink. The accretion radii of the stars are comparable to the binary separation in order to reduce the computational time by neglecting to resolve particle orbits within the binary cavity. We give a resolution study using different sink accretion radii in Section~\ref{sec::resolution}. The disc initially consists of $5\times 10^5$ equal-mass Langragian gas particles and $5\times 10^4$ dust particles that are distributed between the initial inner disc radius $r_{\rm in} = 40\, \rm au \approx 2.5a_{\rm b}$ and the outer disc radius, $r_{\rm out} = 120\, \rm au \approx 7a_{\rm b}$. The circumbinary disc is resolved with a shell-averaged smoothing length per scale height of $\langle h \rangle/H \approx 0.20$. The simulations are modelled for a duration of $t = 1000\, \rm P_{orb}$, where $\rm P_{orb}$ is the binary orbital period. Our standard simulation (run1 from Table~\ref{table::setup}) is extended to $t = 2000\, \rm P_{orb}$. 

The gas surface density profile is initially a power-law distribution given by
 \begin{equation}
     \Sigma(r) = \Sigma_0 \bigg( \frac{r}{r_{\rm in}} \bigg)^{-p},
     \label{eq::sigma}
 \end{equation}
where $\Sigma_0 =  6\times10^{-3}\, \rm g/cm^2$ is the density normalization, $p$ is the power law index, and $r$ is the spherical radius. The density normalization is defined by the total disc mass. The initial disc mass is set to $M_{\rm d} = 0.001M$, where $M$ is the total mass of the binary system, $M=M_1+M_2$.  We ignore the effect of disc self-gravity for the given total disc mass.  We set $p=+3/2$.  We use a locally isothermal equation-of-state given by
\begin{equation}
    c_{\rm s} = c_{\rm s,in} \bigg( \frac{r}{r_{\rm in}} \bigg)^{-q},
\end{equation}
where $c_{\rm s,in}$ is the sound speed at the inner radius.
The disc thickness is scaled with radius as
\begin{equation}
    H = \frac{c_{\rm s}}{\Omega} \propto r^{3/2-q}, 
 \end{equation}
where $\Omega = \sqrt{GM/r^3}$ and $q = +3/4$.  We set an initial gas disc aspect ratio of $H/r = 0.1$ at $r = r_{\rm in}$. The \cite{Shakura1973} viscosity, $\alpha_{\rm SS}$, prescription is given by 
\begin{equation}
    \nu = \alpha_{\rm SS} c_{\rm s} H,
\end{equation}
where $\nu$ is the kinematic viscosity. To calculate $\alpha_{\rm SS}$, we use the prescription in \cite{Lodato2010} given as
\begin{equation}
\alpha_{\rm SS} \approx \frac{\alpha_{\rm AV}}{10}\frac{\langle h \rangle}{H}.
\end{equation}
We take the \cite{Shakura1973} viscosity parameter to be $\alpha_{\rm SS} = 0.01$, which sets the artificial viscosity to $\alpha_{\rm AV} =  2.4$. A value of $\alpha_{\rm AV} = 0.1$ denotes the lower boundary, below which a physical viscosity is not well resolved in SPH, signifying that viscosities smaller than this lower boundary produce disc spreading independently of the value of $\alpha_{\rm AV}$ \cite[see][for details]{Meru2012}. The viscosity prescription also includes a parameter, $\beta_{\rm AV}$, which provides a non-linear term that was originally introduced to prevent particle penetration in high Mach number shocks \cite[e.g.,][]{Monaghan1989}, which is set to $\beta_{\rm AV} = 2.0$. 
The initial disc tilt and binary eccentricity are varied per simulation according to Table~\ref{table::setup}. We simulate two initial disc tilts, $45^\circ$ and $60^\circ$ with a binary eccentricity $e_{\rm b} = 0.8$. Furthermore, we simulate two more binary eccentricities, $e_{\rm b} = 0.3$ and $0.5$ with a $60^\circ$-inclined disc. For each tilt disc and binary eccentricity combination, the initial disc tilt is above the critical tilt for polar alignment. In our follow-up paper, \cite{Smallwood2024} (Paper II), we investigate the effects of the disc surface density, temperature, and viscosity have on the formation of polar dust rings with the application to 99 Her.

The dust particles are initially distributed following the same surface density profile as the gas, with a dust-to-gas mass ratio of $0.01$. Each simulation uses a uniform dust particle size. We simulate average Stokes numbers of $\rm St \approx 15,\, 30,\, 65,\, 100$, which corresponds to particle sizes $s =0.7,1.4,3,4.6\,\rm cm$, respectively.  The Stokes number varies as a function of radius but the dust particle size remains constant. We take the intrinsic grain density to be $3.00\, \rm g/cm^3$.  The initial dust disc aspect ratio is equivalent to the gas disc aspect ratio. 

\subsection{Analysis routine}
In our analysis of the SPH simulations, we break down the disc into $300$ spherical radius bins, spanning from $35\, \rm au$ to $150\, \rm au$. Within each of these bins, we compute various particle characteristics, including surface density, inclination, longitude of ascending node, and eccentricity. An essential factor we emphasize throughout the paper is the dust-to-gas ratio in the midplane. To determine the midplane of a circumbinary disc, we analyse all particles that are situated within half a disc scale height.  

 In each simulation, a dust pileup near the inner edge occurs. This phenomenon could be attributed to physical mechanisms, wherein dust particles drifting inward are impeded at the gas inner edge, preventing further drift. Alternatively, this could be an artifact of numerical resolution effects; as the gas resolution diminishes at the inner edge, the locally better-resolved dust appears amplified \citep{Ayliffe2012}. The warp and alignment physics (differential precession) can be ruled out as an origin since evidence shows that this effect occurs and potentially strengthens in flat, planar discs as well \citep[Figure 4 in][]{Aly2021}. Resonances induced by the binary may also contribute, particularly as the dust pileup intensifies with higher Stokes numbers. Therefore, this phenomenon cannot be attributed to a 'traffic jam' effect; however, its exact origins still need to be clarified through further investigation.

\begin{table}
	\centering
	\caption{The setup of the gas and dust SPH simulations that includes an initial circumbinary disc. The table lists the initial binary eccentricity, $e_{\rm b}$, initial Stokes number, $\rm St$, and initial tilt of the disc, $i_{\rm d}$
 }
	\begin{tabular}{lccc} 
		\hline
	    Model & $e_{\rm b}$ & $\rm St$ & $i_{\rm d}$\\
		\hline
		\hline
            run1 & $0.8$  & $65$ & $60$ \\
		run2 & $0.8$  & $15$ & $60$ \\
            run3 & $0.8$  & $30$ & $60$  \\
            run4 & $0.8$  & $100$ & $60$\\
		run5 & $0.3$  & $65$ & $60$ \\
		run6 & $0.5$  & $65$ & $60$  \\
		run7 & $0.8$  & $65$ & $45$ \\
		\hline
	\end{tabular}
    \label{table::setup}
\end{table}

\section{Hydrodynamical Results}
\label{sec::hydro_results}


\begin{figure*} 
\centering
\includegraphics[width=2\columnwidth]{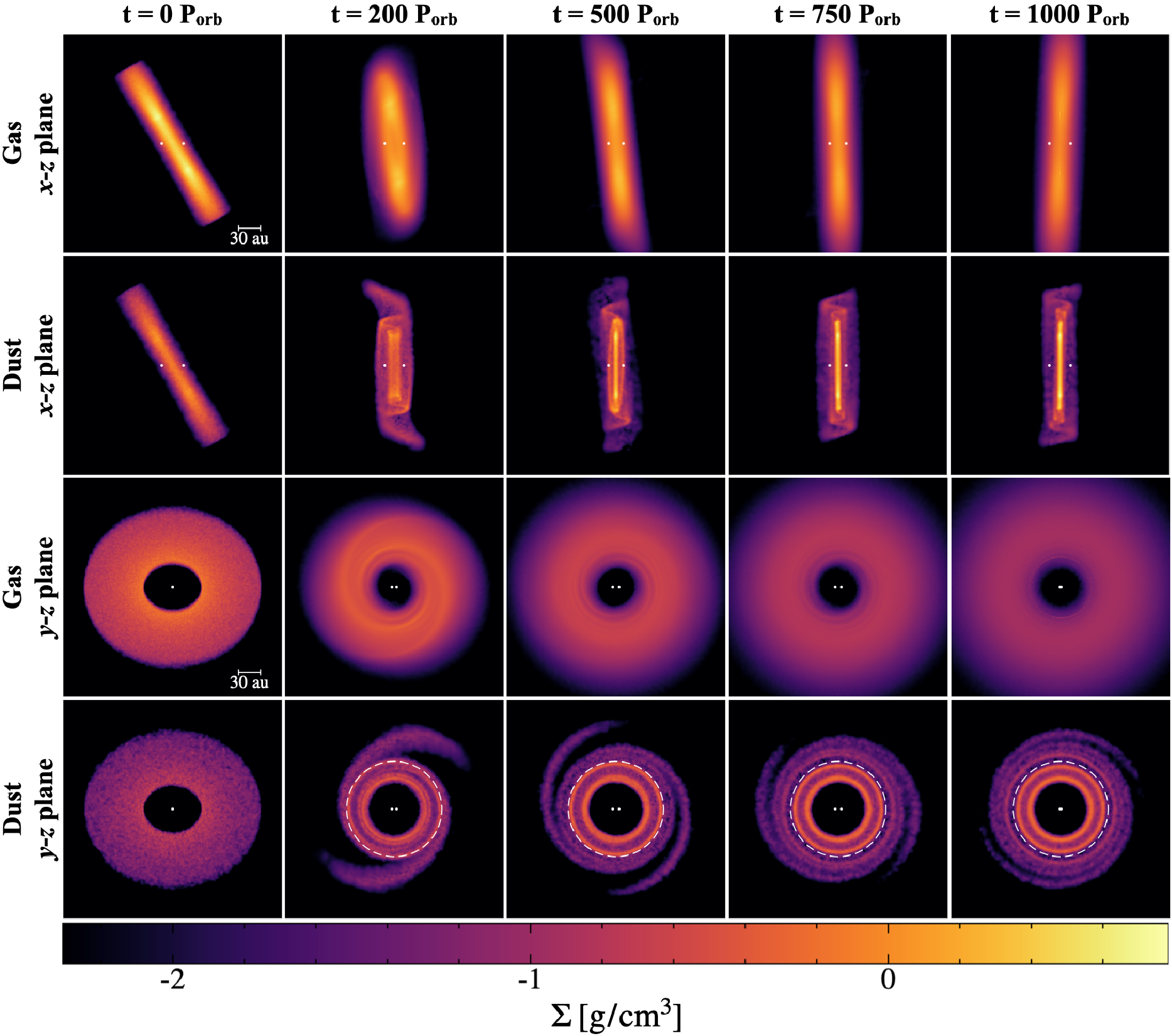}
\centering
\caption{The structure and evolution of a highly inclined dusty circumbinary disc with $\rm St \sim 65$ (run1). The top two rows show the gas and dust disc structure viewing in the $x$-$z$ plane, while the bottom two rows are viewing the disc in the $y$-$z$ plane. The columns show select times up to $1000\, \rm P_{orb}$.  Initially both the gas and dust are misaligned but align to a polar configuration, with dust traffic jams producing dust rings. The location of the initial dust ring is showed by the white-dotted circle. }
\label{fig::splash_dust}
\end{figure*}

\begin{figure*} 
\centering
\includegraphics[width=\columnwidth]{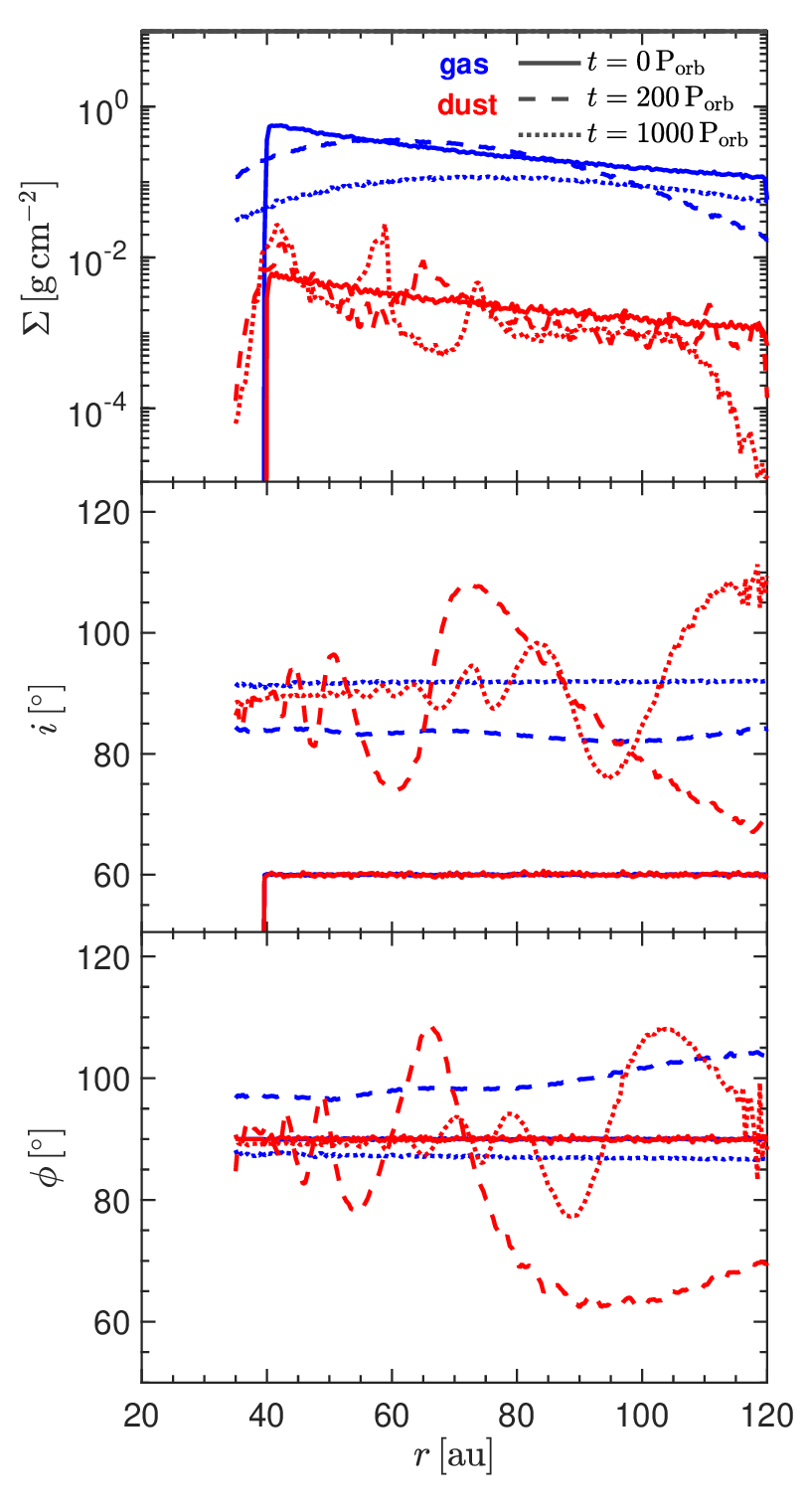}
\includegraphics[width=\columnwidth]{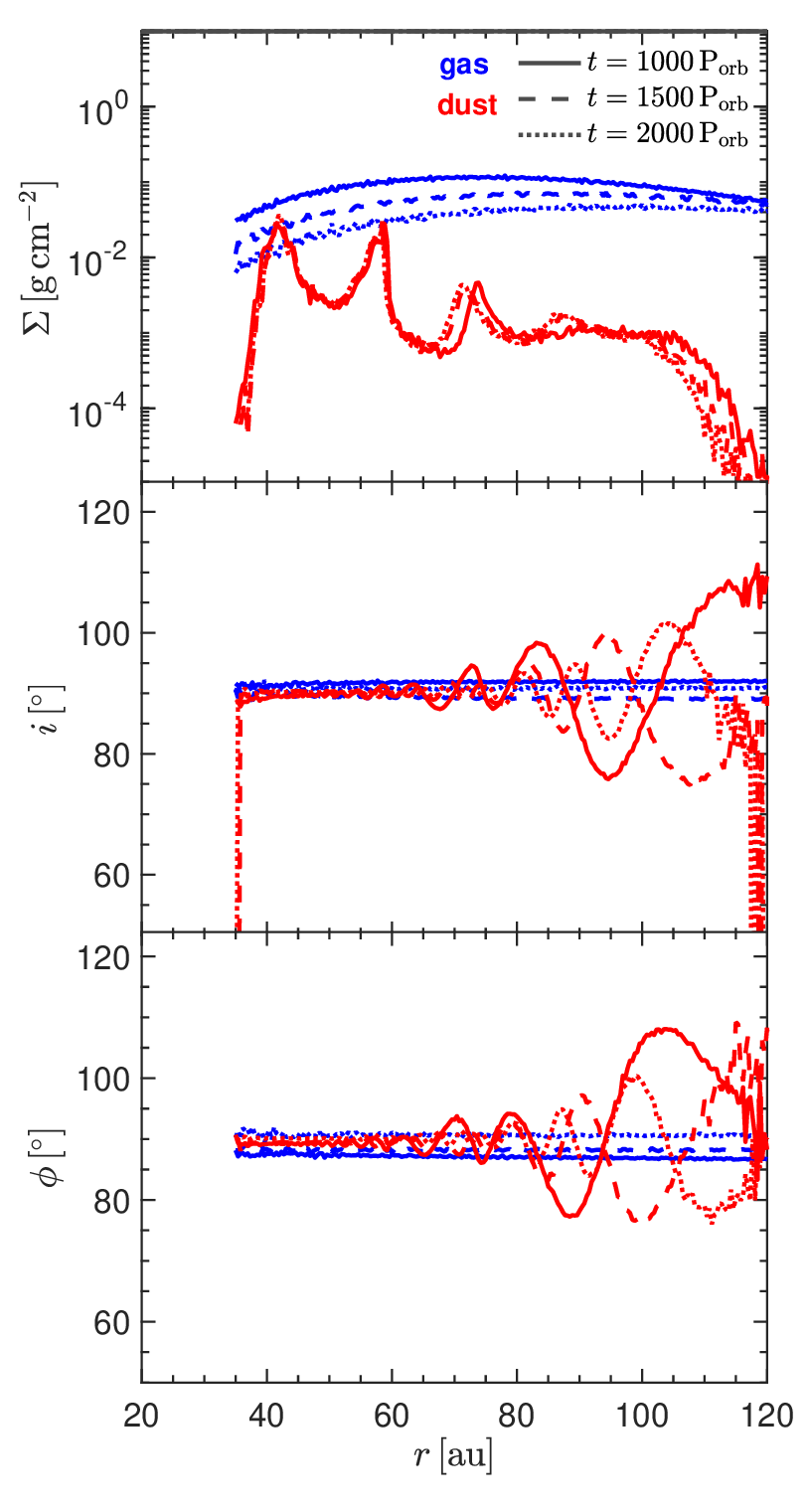}
\centering
\caption{The surface density $\Sigma$ (top panel), tilt $i$ (middle panel), and longitude of ascending node $\phi$ (bottom panel) as a function of disc radius $r$ for our standard disc simulation with $\rm St \sim 65$ (run1). The blue lines correspond to the gas, while the red corresponds to the dust.  The left plot shows the times $t = 0\, \rm P_{orb}$ (solid line), $t = 200\, \rm P_{orb}$ (dashed line), and $t = 1000\, \rm P_{orb}$ (dotted line). The right plot shows the times $t = 1000\, \rm P_{orb}$ (solid line), $t = 1500\, \rm P_{orb}$ (dashed line), and $t = 2000\, \rm P_{orb}$ (dotted line). The differential precession between the gas and dust leads to the formation of dust traffic jams in the circumbinary disc.}
\label{fig::disc_params_St65}
\end{figure*}

\begin{figure*} 
\centering
\includegraphics[width=\columnwidth]{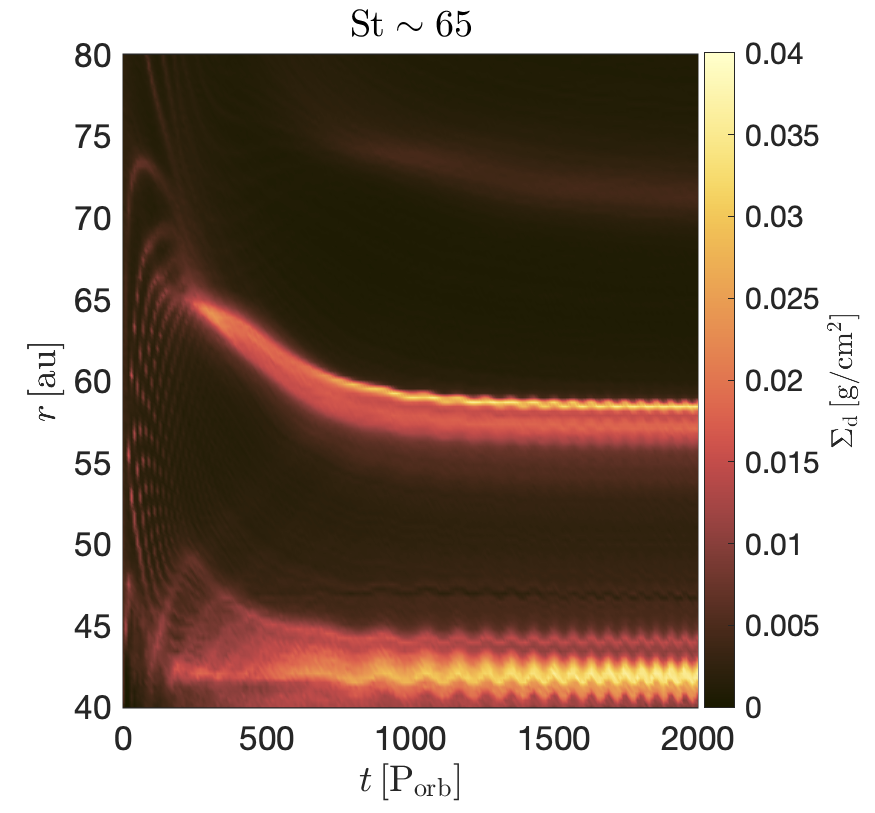}
\includegraphics[width=\columnwidth]{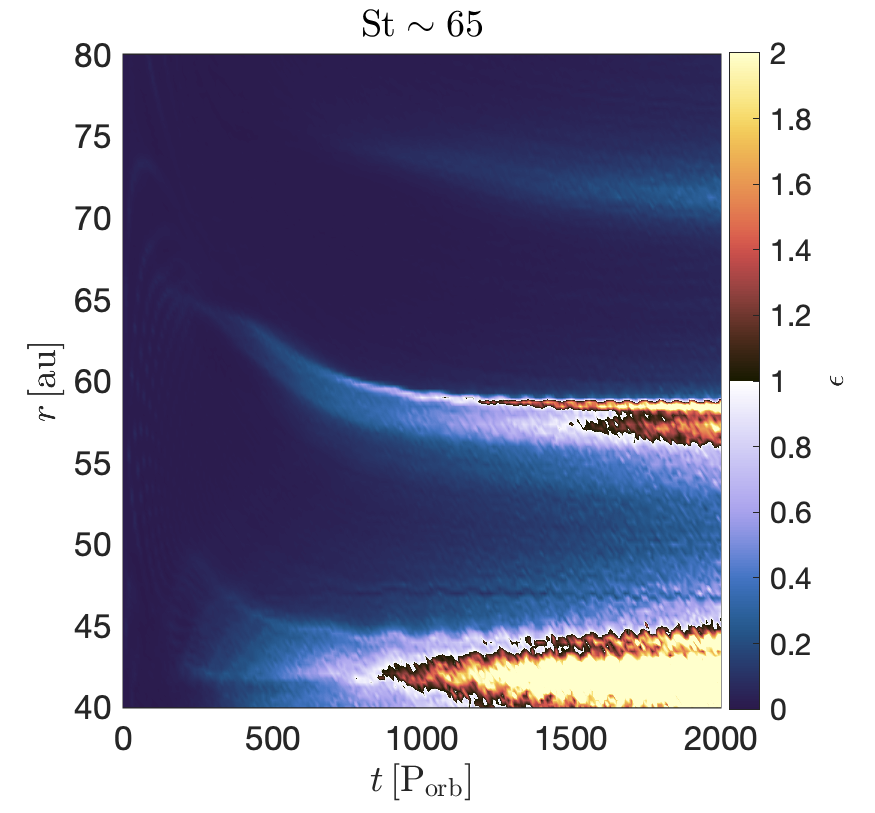}
\centering
\caption{ {\it Left panel}: the circumbinary disc dust surface density, $\Sigma_{\rm d}$, as a function disc radius, $r$, and time in binary orbital periods, $\rm P_{orb}$, for our standard model with $\rm St \sim 65$ (run1). {\it Right panel}: the circumbinary disc midplane dust-to-gas ratio, $\epsilon = \rho_{\rm d}/\rho_{\rm g}$.}
\label{fig::sigma_extend}
\end{figure*}

\subsection{Standard Simulation}
An initially inclined gaseous circumbinary disc may undergo polar alignment, where the disc angular momentum will precess about, and eventually align with, the binary eccentricity vector \citep{Martinlubow2017,Smallwood2020}. During the alignment process, the disc tilt will oscillate due to the binary torque \citep{Smallwood2020}. Once the disc is nearly in a polar configuration, the tilt oscillations decrease. It is unclear how the dust reacts during polar alignment, and thus we first showcase our standard simulation ( run1 from Table~\ref{table::setup}) of an initially misaligned circumbinary disc with gas and dust components  analysed up to $2000\, \rm P_{orb}$.

The dust grains in the standard simulation have $\rm St \sim 65$. We visualise our standard simulation in Fig.~\ref{fig::splash_dust}  up to $1000\, \rm P_{orb}$.  The disc structure at $1000\, \rm P_{orb}$ is similar to $2000\, \rm P_{orb}$ (see Fig.~\ref{fig::disc_params_St65}). However, we show the face-on view of the disc at $2000\, \rm P_{orb}$ in Appendix~\ref{app::extended}. The top two rows in Fig.~\ref{fig::splash_dust} show the gas and dust disc structures in the $x$--$z$ plane, with the colour denoting the surface density. The disc is initially inclined by $60^\circ$ and aligns to a polar configuration. At $1000\, \rm P_{orb}$, the gaseous disc is perpendicular to the binary orbital plane. At the same time, the inner regions of the dust disc are polar aligned, while the outer portions are still undergoing alignment. The reason being that at the inner radii, dust experiences stronger binary torque, and the Stokes number is smaller (it increases with radius since it is inversely proportional to the density). The bottom two rows show the gas and dust disc structures in the $y$--$z$ plane. As the disc aligns polar, the disc becomes face-on in the $y$--$z$ plane. The dust traffic jam appears within the dust disc at $t=200\, \rm P_{orb}$.  The dust traffic jam will eventually evolve into a dense dust ring (see Appendix~\ref{app::dustring}). The dashed-white circle marks the initial location of the dust ring. As the simulation progresses, the dust ring begins to drift inward (see times $500\, \rm P_{orb}$ and $750\, \rm P_{orb}$). At $1000\, \rm P_{orb}$, the original dust ring is interior to its initial location (the white circle). Also, at this time, a secondary dust ring is beginning to form exterior to the white circle.

We examine more closely the difference in dynamics between gas and dust for our standard simulation. The left panel in Fig.~\ref{fig::disc_params_St65} shows the disc surface density (upper panel), tilt (middle panel), and longitude of the ascending node (lower panel) as a function of disc radius at $t = 0 \, \rm P_{orb}$, $200\, \rm P_{orb}$, and $1000\, \rm P_{orb}$.  There are no pressure bumps within the disc, allowing the gas profile to be largely smooth in radius. The gas surface density evolves slowly in time, where the inner parts of the disc flow through the binary cavity and accrete onto the binary components and the outer parts of the disc viscously spread outwards. At $200\, \rm P_{orb}$ there is a peak in the dust surface density corresponding to the first dust traffic jam that is located at $r \sim 65\, \rm au$. At $t=1000\, \rm P_{orb}$, the initial dust ring drifts inward to $\sim 55\, \rm au$. There is a second dust traffic jam that is beginning to form at $r \sim 75\, \rm au$. 
Both the gas and dust initially have a disc tilt of $60^\circ$. The entire gas disc eventually evolves close to a polar configuration, $i \sim 90^\circ$. At $t = 200\, \rm P_{orb}$, the dust is strongly warped, especially when the disc radius increases. At $t = 1000\, \rm P_{orb}$, the dust in the inner regions of the disc, $r <65\, \rm au$, are in a nearly polar state, while the dust at $r > 65\, \rm au$ are still warped.  At $200\, \rm P_{orb}$, there is a large difference between the precession of gas and dust located at $r \sim 65\, \rm au$, which corresponds to the location of the first dust traffic jam (see top panel). At  $t = 1000\, \rm P_{orb}$, there is still a difference between the gas and dust precession rates at $r > 65\, \rm au$.


The right panel in Fig.~\ref{fig::disc_params_St65} shows the disc surface density (upper panel), tilt (middle panel), and longitude of the ascending node (lower panel) as a function of disc radius at $t = 1000 \, \rm P_{orb}$, $1500\, \rm P_{orb}$, and $2000\, \rm P_{orb}$. The density of the initial dust ring at $t = 1000\, \rm P_{orb}$ is consistent with the density at $t = 2000\, \rm P_{orb}$. During this time, the secondary dust ring drifts inward and a third dust ring begins to form at $r \sim 90\, \rm au$. As time goes on, the surface density of the gas decreasing as material is accreted by the binary.  At $t = 2000\, \rm au$, the whole gas disc is aligned in a nearly polar state. At the same time, the dust disc at $r \lesssim 80\, \rm au$ is aligned in a polar state, while the dust at $r \gtrsim 80\, \rm au$  is still misaligned to the gas. Therefore, the engine to produce dust rings is still operating within the outer regions of the disc.  At $1500\, \rm P_{orb}$ and $2000\, \rm P_{orb}$, there is still a difference between the precession of gas and dust located at $r \gtrsim 80\, \rm au$, which means dust rings will continue to be produced until both components (gas and dust) are polar aligned.

The left panel of Fig.~\ref{fig::sigma_extend} shows the azimuthally-averaged disc surface density as a function of disc radius and time. The dust rings drifts inwards and reaches a quasi-steady state beginning at $t \sim 1000\, \rm P_{orb}$. Beyond $1000\, \rm P_{orb}$, the dust ring is stable with a constant surface density. A secondary dust traffic jam begins to form at $t \sim 750\, \rm P_{orb}$ at $r \sim 75\, \rm au$ and also begins to drift inward. As the simulation progresses, the Stokes number of the dust grains increases since the gas surface density decreases due to accretion onto the binary. Eventually, the dust ring does not feel the drag of the gas, and the radial location of the ring remains at a given radius (see Appendix~\ref{app:ring_location}), along with oscillations due to the interaction with the binary.

\begin{figure} 
\centering
\includegraphics[width=\columnwidth]{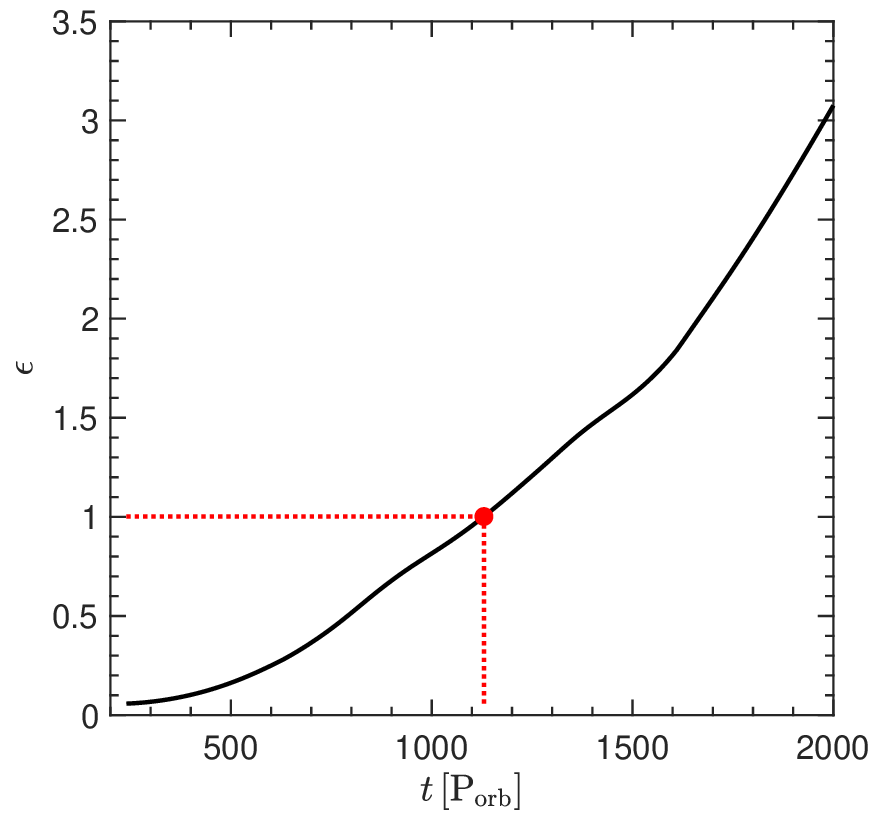}
\centering
\caption{ We trace the midplane dust-to-gas ratio, $\epsilon$, from Fig.~\ref{fig::sigma_extend} along the initial dust ring as a function of time in binary orbital periods, $\rm P_{orb}$. The red dot indicates when $\epsilon \simeq 1$, which occurs at $t \sim  1130\, \rm P_{orb}$. As time progresses, $\epsilon$ exhibits a nearly exponential growth, approaching $\epsilon \approx 3$ by the conclusion of the simulation.} 
\label{fig::mid_time}
\end{figure}

The right panel of Fig.~\ref{fig::sigma_extend} shows the azimuthally-averaged dust-to-gas ratio in the midplane, $\epsilon = \rho_{\rm d}/\rho_{\rm g}$, as a function of disc radius and time. As the simulation progresses, $\epsilon$ continues to grow within the dust rings. Figure~\ref{fig::mid_time} shows the midplane dust-to-gas ratio, $\epsilon$, as a function of binary orbital periods. The dust-to-gas ratio is traced along the initial dust ring from Fig.~\ref{fig::sigma_extend}. Over the course of the simulation, $\epsilon$ exhibits a nearly exponential growth. At $t \sim  1130\, \rm P_{orb}$, $\epsilon$ reaches unity. By the conclusion of the simulation,  $\epsilon\sim 3$. In numerical simulations, when the dust-to-gas ratio surpasses unity in the midplane, it can induce the streaming instability to trigger significant clumping \cite[e.g.,][]{Johansen2009,Schafer2017,Nesvorny2019,Li2021}.

\begin{figure*} 
\centering
\includegraphics[width=\columnwidth]{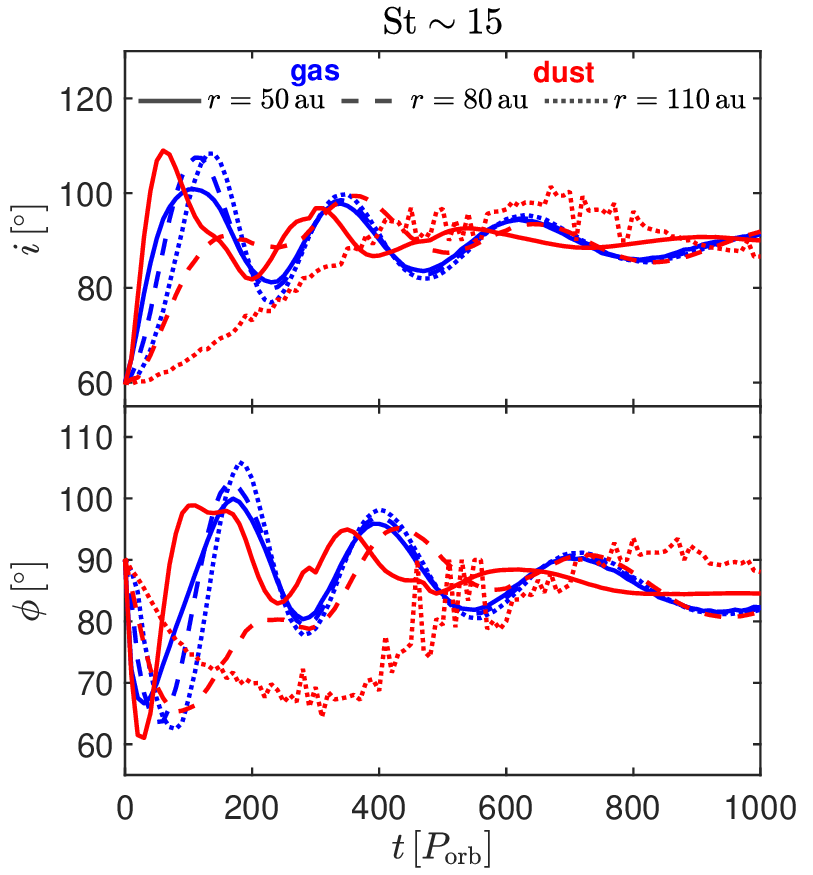}
\includegraphics[width=\columnwidth]{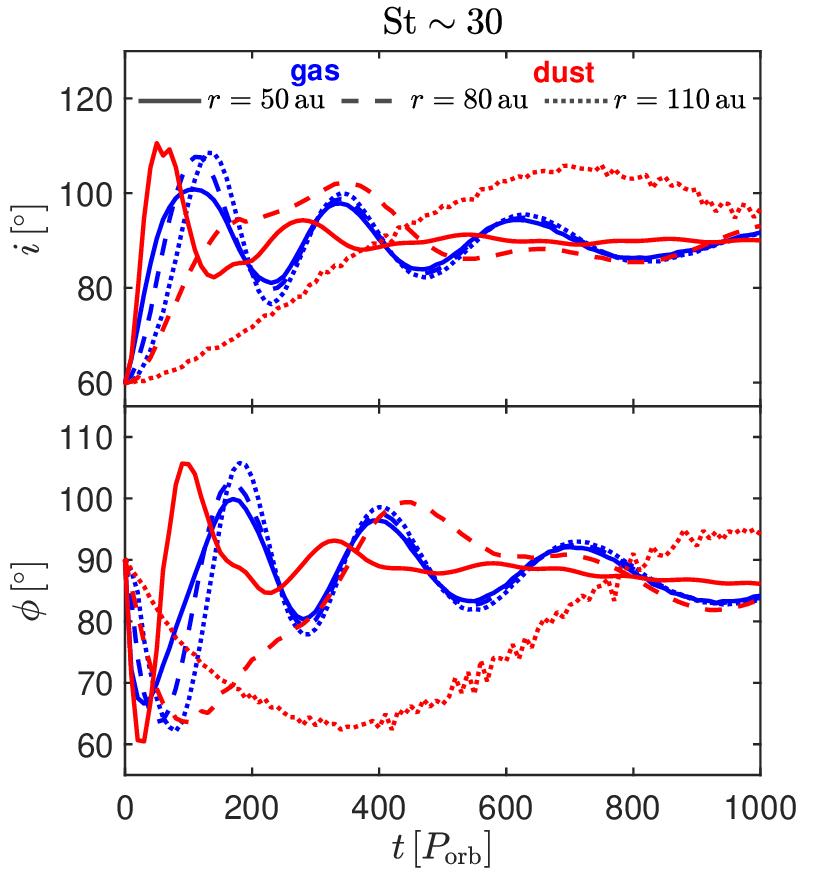}
\includegraphics[width=\columnwidth]{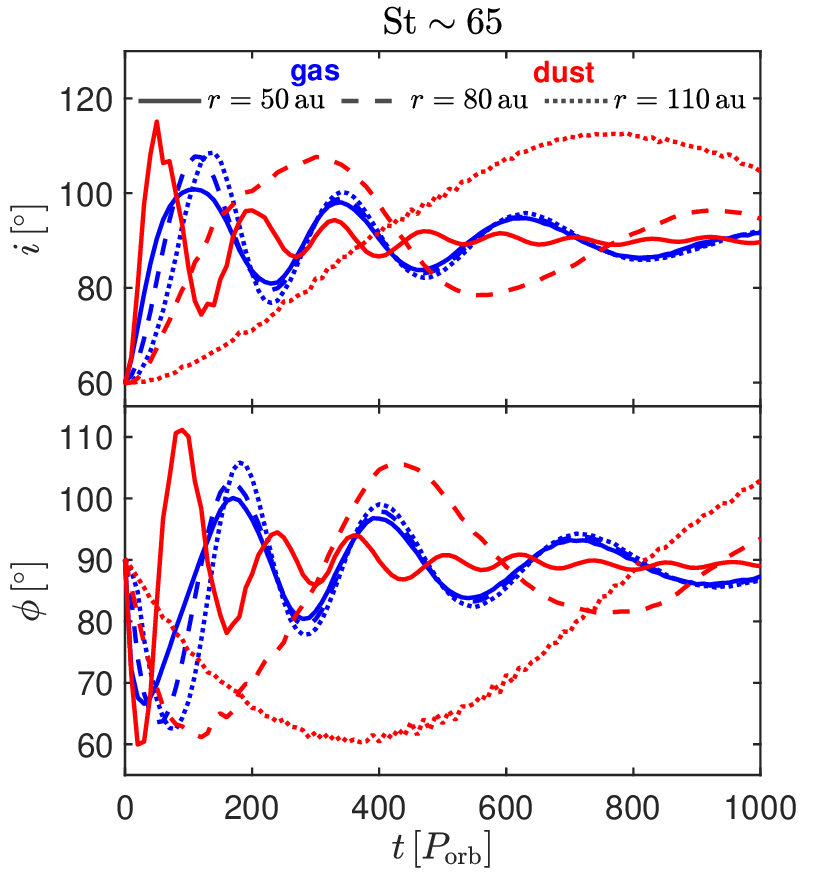}
\includegraphics[width=\columnwidth]{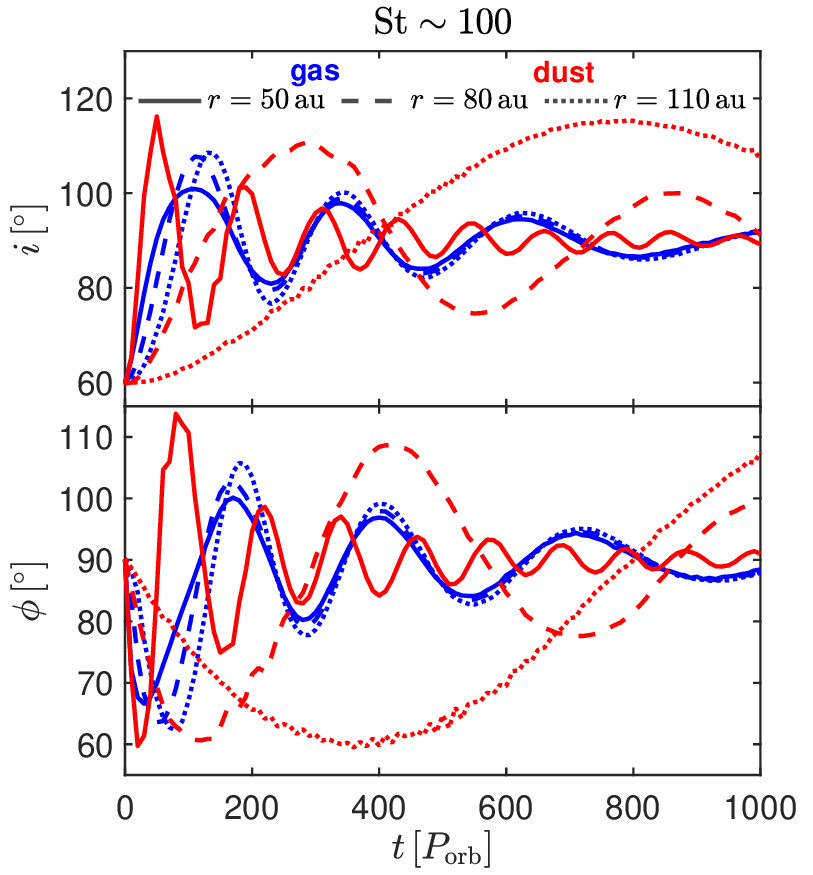}
\centering
\caption{The evolution of the disc tilt $i$ and longitude of the ascending node $\phi$ as a function of time in units of the binary orbital period $P_{\rm orb}$ for different initial Stokes number, $\rm St$. Top left panel: $\rm St \sim 15$ (run2). Top right panel: $\rm St \sim 30$ (run3). Bottom left panel: $\rm St \sim 65$ (our standard model, run1). Bottom right panel: $\rm St \sim 100$ (run4). The gas is given by the blue curves, and the dust is given by the red curves. We probe the disc at $50\, \rm au$ (solid), $80\, \rm au$ (dashed), and $110\, \rm au$ (dotted).}
\label{fig::disc_params_St}
\end{figure*}

\begin{figure*} 
\centering
\includegraphics[width=\columnwidth]{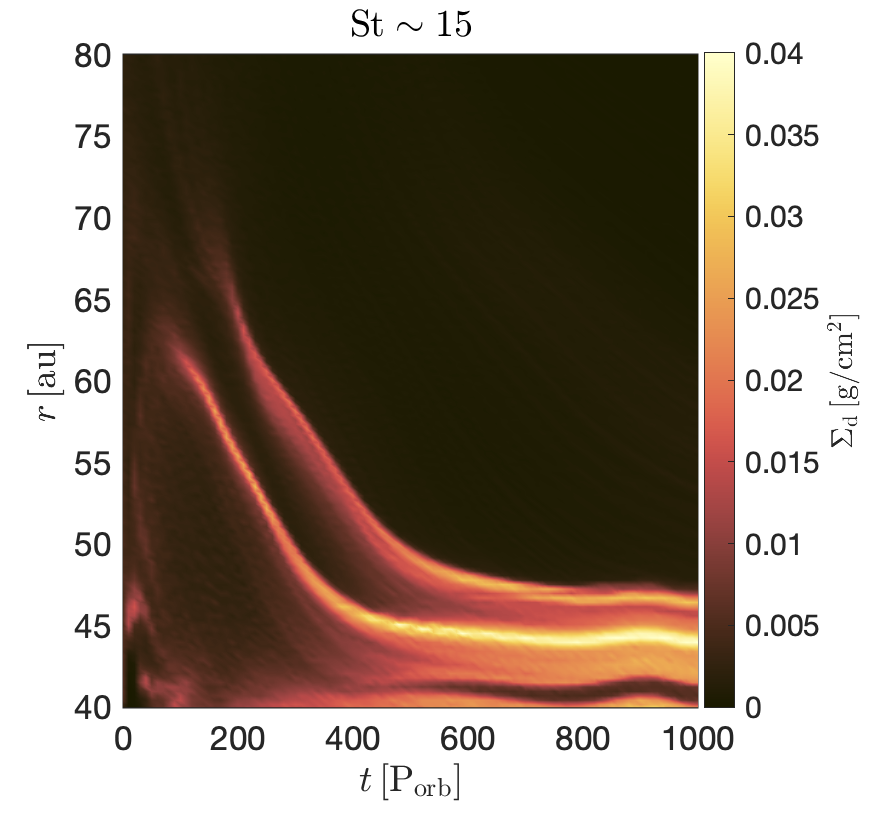}
\includegraphics[width=\columnwidth]{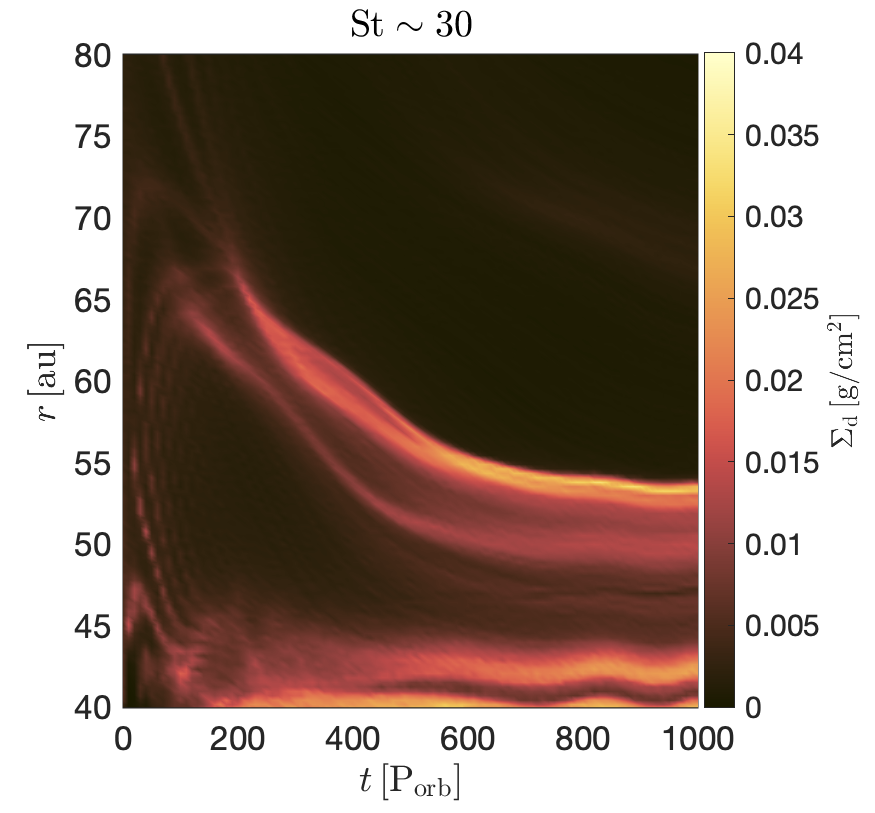}
\includegraphics[width=\columnwidth]{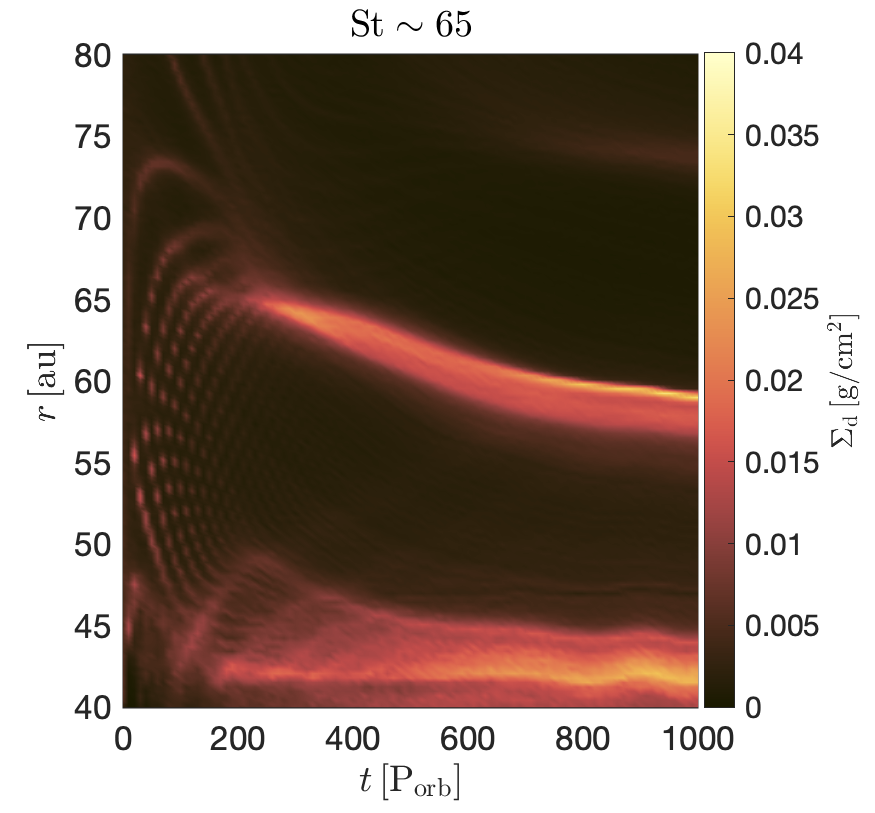}
\includegraphics[width=\columnwidth]{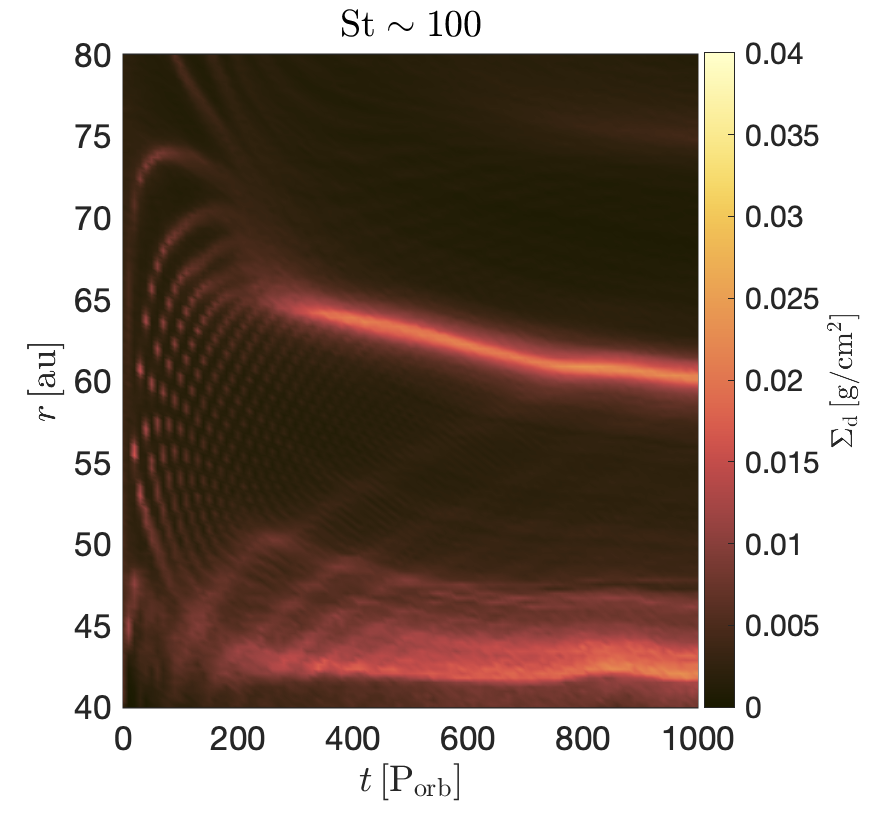}
\centering
\caption{The circumbinary disc dust surface density, $\Sigma_{\rm d}$, as a function disc radius, $r$, and time in binary orbital periods, $\rm P_{orb}$, for different initial Stokes number, $\rm St$. Top left panel: $\rm St \sim 15$ (run2). Top right panel: $\rm St \sim 30$ (run3). Bottom left panel: $\rm St \sim 65$ (our standard model, run1).  The panel is the same as in Fig.~\ref{fig::sigma_extend} but truncated to $1000\, \rm P_{orb}$ for comparison. Bottom right panel: $\rm St \sim 100$ (run4). The colour map denotes the dust surface density. In each model a dust traffic jam is produced, and drifts inward.}
\label{fig::sigma_St}
\end{figure*}

\subsection{Varying Stokes number}
We analyze the simulations with different Stokes number, $\rm St \sim 15$ (run2), $30$ (run3), $65$ (run1), and $100$ (run4). We  first analyse the evolution of the gas and dust as the circumbinary disc aligns to a polar state. Figure~\ref{fig::disc_params_St} shows the disc tilt and longitude of the ascending node as a function of time  for $\rm St \sim 15$ (top-left panel), $\rm St \sim 30$ (top-right panel), $\rm St \sim 65$ (bottom-left panel), and $\rm St \sim 100$ (bottom-right panel). The gas is given by the blue lines, while the dust is given by the red lines. We probe the disc at three different radial distances, $r = 50$, $80$, $110\, \rm au$. The distance $r = 50\, \rm au$, is near the initial inner disc edge, while $r = 110\, \rm au$ is near the initial outer disc edge. In each Stokes number case, the gas disc is undergoing a smooth alignment to a polar configuration, where the inner parts of the disc align on a faster timescale than the outer parts of the disc.  In each simulation, the disc does not break during the alignment process because the disc precesses nearly as a rigid body since the radial communication timescale is shorter than the precession timescale \cite[e.g.,][]{papaloizou1995,Larwood1997}. During the alignment process, the gas disc undergoes tilt oscillations driven by the binary \cite[e.g.,][]{Smallwood2019,Smallwood2020}. Although within $\pm5^{\circ}$, the gas disc has not fully aligned polar within the simulation time domain of $1000\, \rm P_{\rm orb}$, which was selected for computational reasons since we are modeling the interaction of gas and dust particles. However, a polar-aligning circumbinary disc is expected to fully align polar well before the lifetime of the gas disc \cite[see Section~5 in][]{Smallwood2020}.

\begin{figure*} 
\centering
\includegraphics[width=\columnwidth]{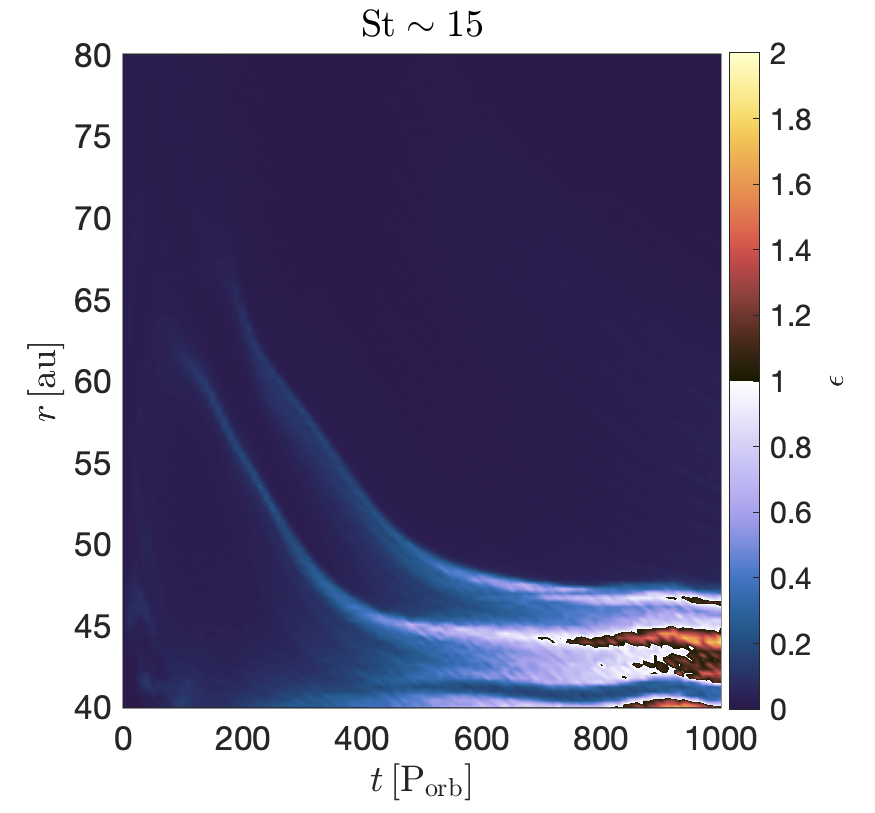}
\includegraphics[width=\columnwidth]{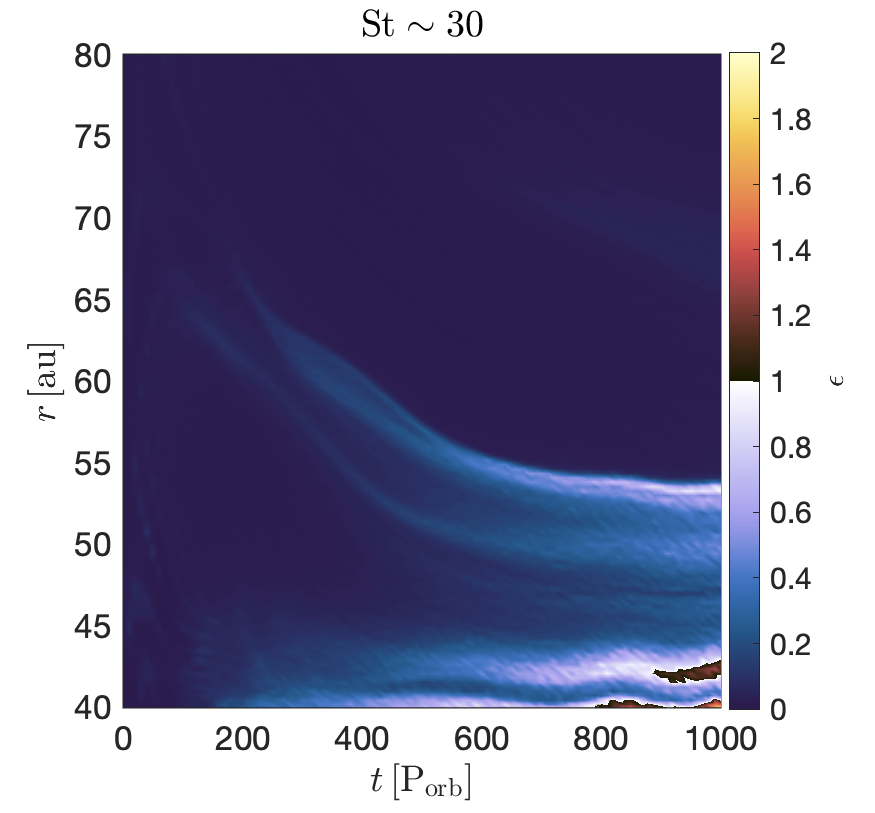}
\includegraphics[width=\columnwidth]{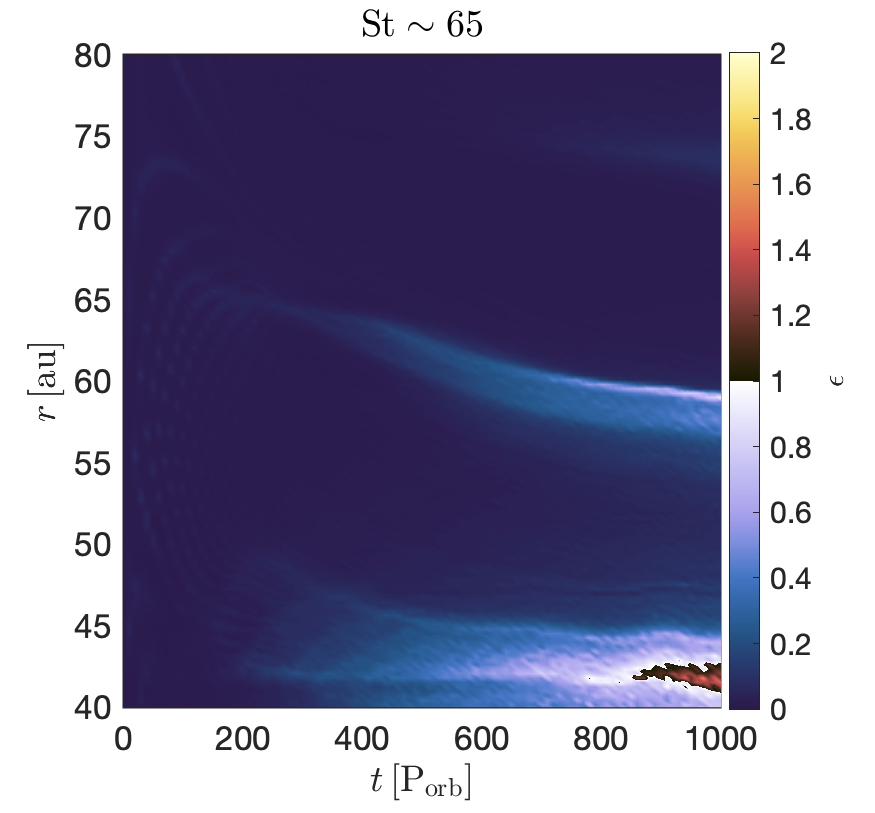}
\includegraphics[width=\columnwidth]{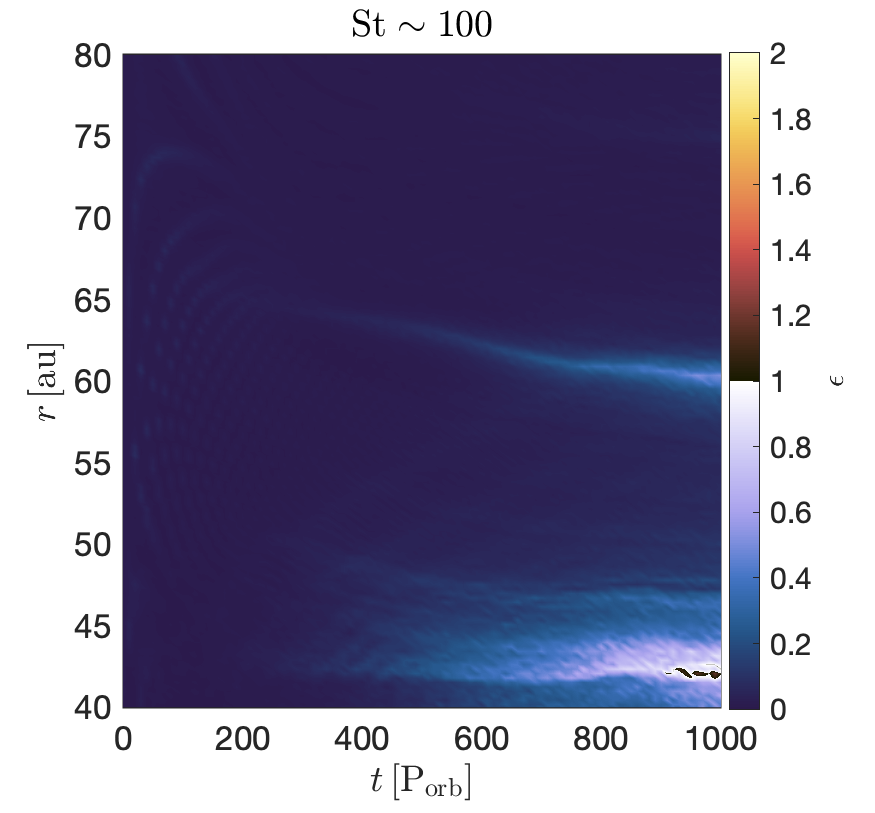}
\centering
\caption{The circumbinary disc midplane dust-to-gas ratio, $\epsilon$, for various Stokes number, $\rm St$, as a function of disc radius, $r$, and time in binary orbital periods, $\rm P_{orb}$. Top-left panel: $\rm St \sim 15$ (run2). Top-right panel: $\rm St \sim 30$ (run3). Bottom-left panel: $\rm St \sim 65$ (run1).  The panel is the same as in Fig.~\ref{fig::sigma_extend} but truncated to $1000\, \rm P_{orb}$ for comparison. Bottom-right panel: $\rm St \sim 100$ (run4).  As the dust ring evolves in time,  $\epsilon$ increases towards unity which can trigger the streaming instability. }
\label{fig::mid_St}
\end{figure*}

The dust evolves quite differently than the gas in Fig.~\ref{fig::disc_params_St}. The dust undergoes tilt oscillations like the gas, but the alignment timescale for the dust is much longer for dust located in the outer parts of the disc. In each case, the dust located at $50\, \rm au$ nearly aligns to a polar configuration in each different Stokes number simulation.  As the Stokes number increases, the period of the tilt oscillation in the dust gets smaller, and the alignment timescale of the dust increases. The precession period of a test particle increases with radius. The gas disc precession period is equal to a test particle period at intermediate disc radii. The tilt oscillations in the gas in the inner parts of the disc have a larger period than a test particle at the same radius. The dust at lower Stokes number is more influenced by the gas and thus has a larger tilt oscillation period than a test particle. The dust at high Stokes number behaves more like a test particle, and therefore has a longer tilt oscillation period than dust at lower Stokes number. The tilt oscillation for the outer region of the disc has not completed one full oscillation within $1000\, \rm P_{orb}$. Also, the amplitude of this oscillation increases as the  Stokes number increases. In each Stokes number case, there is clear difference in the precession between the gas and dust seen by the evolution of the tilt and longitude of the ascending node. This difference is the engine that generates the dust traffic jams \cite[e.g.,][]{Aly2021}. 


Figure~\ref{fig::sigma_St} shows the azimuthally-averaged disc surface density as a function of disc radius and time for $\rm St \sim 15$ (top-left panel), $\rm St \sim 30$ (top-right panel), $\rm St \sim 65$ (bottom-left panel), and $\rm St \sim 100$ (bottom-right panel). In each case, dust traffic jams occur due to the difference in precession between the gas and dust \cite[e.g.,][]{Aly2021}. For $\rm St \sim 15$, an initial dust ring forms at a radial distance $r\sim 63\, \rm au$ at $t \sim 150\, \rm P_{orb}$. A secondary dust ring forms at $r\sim 72\, \rm au$ at $t \sim 200\, \rm P_{orb}$. Together, the two dust rings drift inward and reach a steady-state radius when $t \gtrsim 600\, \rm P_{orb}$. Eventually, the presence of an outer, secondary dust ring will suffocate the growth of the initial, inner dust ring \cite[e.g.,][]{Aly2021}. For $\rm St \sim 30$, a two-dust ring structure is also produced. The initial, inner dust ring forms at $r\sim 64\, \rm au$ at $t \sim 150\, \rm P_{orb}$, and the secondary, outer dust rings forms at $r\sim 73\, \rm au$ at $t \sim 200\, \rm P_{orb}$. Similar to the $\rm St\sim 15$ case, the dust rings drift inward and eventually stall at $r \sim 55\, \rm au$.  For $\rm St \sim 65$, our standard model, the initial dust traffic jam occurs at $r \sim 65\, \rm au$ at $t \sim 250\, \rm P_{orb}$. The dust rings drift inwards at a slower rate compared to the smaller Stokes number models. A secondary dust traffic jam begins to form at $t \sim 750\, \rm P_{orb}$ at $r \sim 75\, \rm au$. For the simulation with the highest Stokes number, $\rm St \sim 100$, an initial dust traffic jam occurs at $r \sim 65\, \rm au$ at $t \sim 250\, \rm P_{orb}$ (similar to $\rm St \sim 65$). However, unlike the $\rm St \sim 65$ model, the dust ring drifts inward at a slower rate.  A secondary dust traffic jams begins to emerge at $t \sim 750\, \rm P_{orb}$ at a radius $r \sim 75\, \rm au$. In general, the initial position of the dust traffic jams is the same for every Stokes number. Then, the traffic jams evolve differently depending on the Stokes number. As the Stokes number increases, the initial dust ring forms further out in the disc at later times. The inward drift of the dust rings is faster for lower Stokes number,  since such particles experience stronger gas drag.

\begin{figure} 
\centering
\includegraphics[width=\columnwidth]{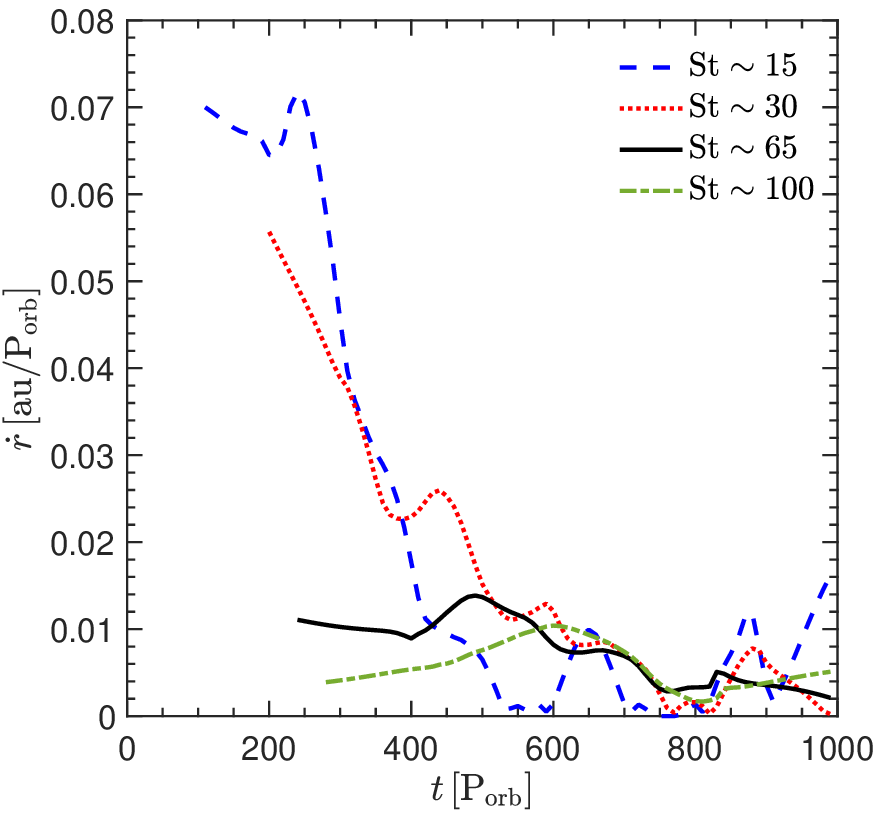}
\centering
\caption{The drift rates, $\dot{r}$, of the initial dust rings as a function of time in binary orbital periods, $\rm P_{orb}$, for the four different Stokes number simulations: $\rm St\sim 15$ (blue-dashed, run2), $\rm St\sim 30$ (red-dotted, run3), $\rm St\sim 65$ (black-solid, run1), and $\rm St\sim 100$ (green-dash-dotted, run4). The lower Stokes number simulations will have a higher dust ring drift rate. 
}
\label{fig::drift_St}
\end{figure}

\begin{figure} 
\centering
\includegraphics[width=\columnwidth]{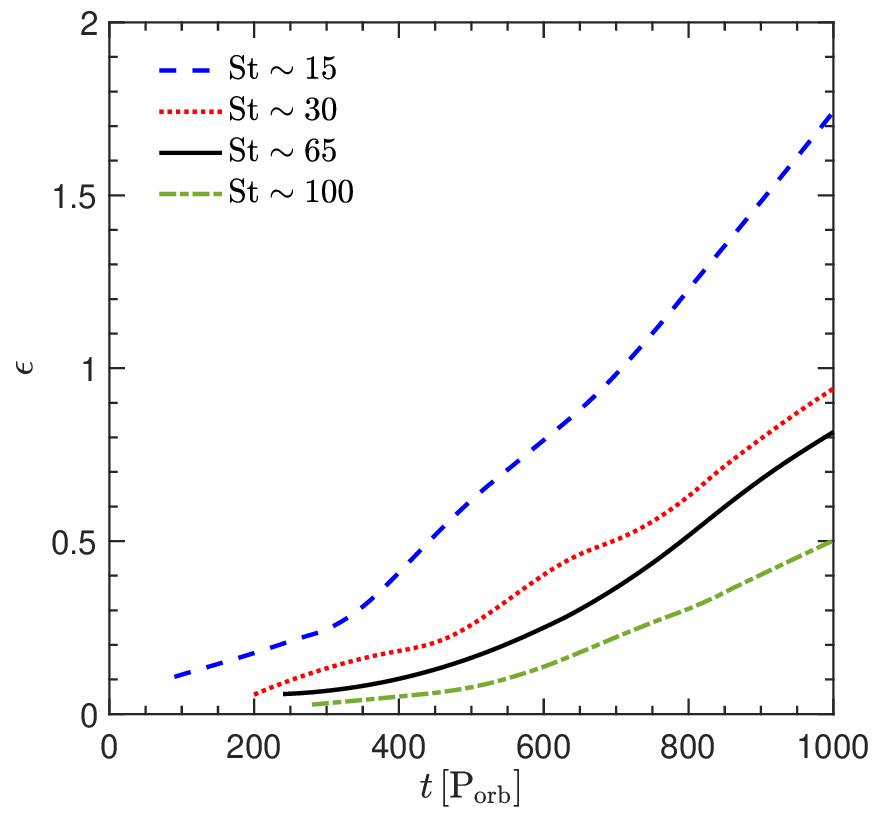}
\centering
\caption{We trace the midplane dust-to-gas ratio, $\epsilon$, from Fig.~\ref{fig::mid_St} along the initial dust ring as a function of time in binary orbital periods, $\rm P_{orb}$, for each Stokes number simulation, $\rm St\sim 15$ (blue-dashed, run2),  $\rm St\sim 30$ (red-dotted, run3),  $\rm St\sim 65$ (black-solid, run1), and  $\rm St\sim 100$ (green-dash-dotted, run4). The black curve ($\rm St \sim 65$) is the same as in Fig.~\ref{fig::mid_time} but truncated to $1000\, \rm P_{orb}$ for comparison. As time progresses, $\epsilon$ exhibits a nearly exponential growth, approaching $\epsilon \approx 1$.}

\label{fig::mid_time_st}
\end{figure}


\begin{figure*} 
\centering
\includegraphics[width=\columnwidth]{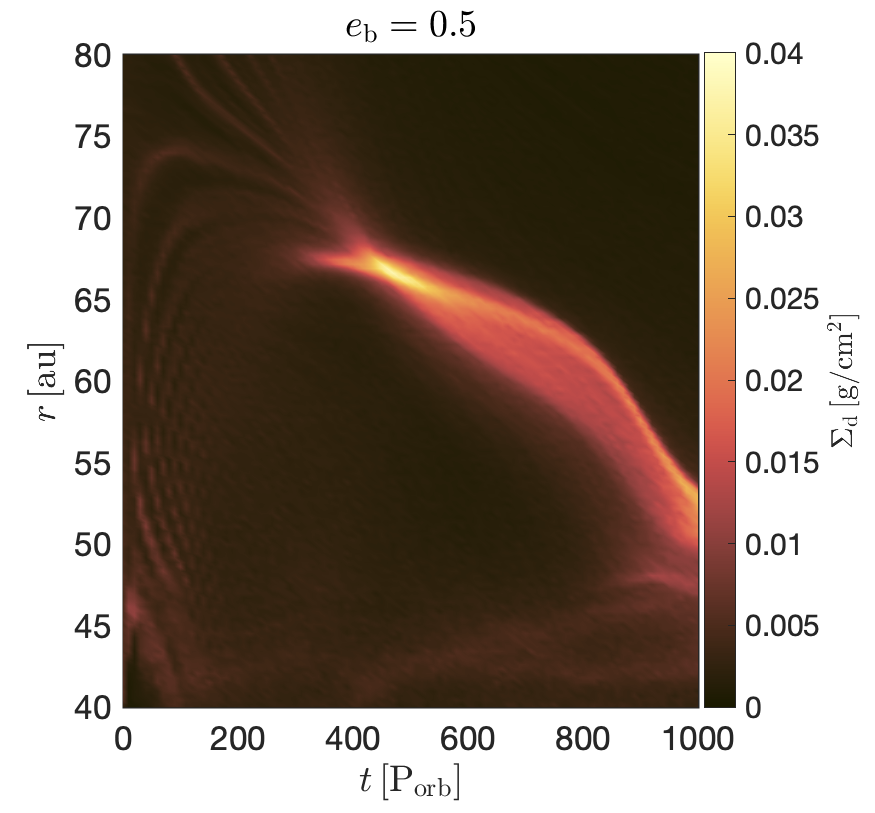}
\includegraphics[width=\columnwidth]{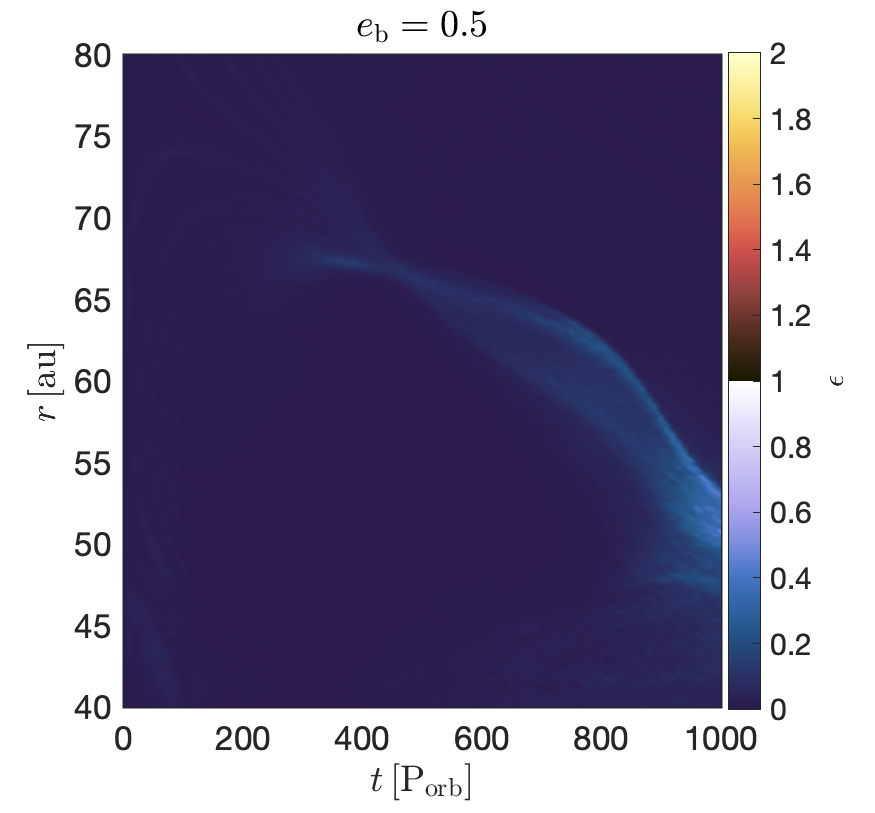}
\centering
\caption{{\it Left panel}: the circumbinary disc dust surface density, $\Sigma_{\rm d}$, as a function disc radius, $r$, and time in binary orbital periods, $\rm P_{orb}$, for a binary eccentricity $e_{\rm b} = 0.5$ (run6). {\it Right panel}: the circumbinary disc midplane dust-to-gas ratio, $\rho_{\rm d}/\rho_{\rm g}$.}
\label{fig::sigma_ecc}
\end{figure*}

\begin{figure*} 
\centering
\includegraphics[width=\columnwidth]{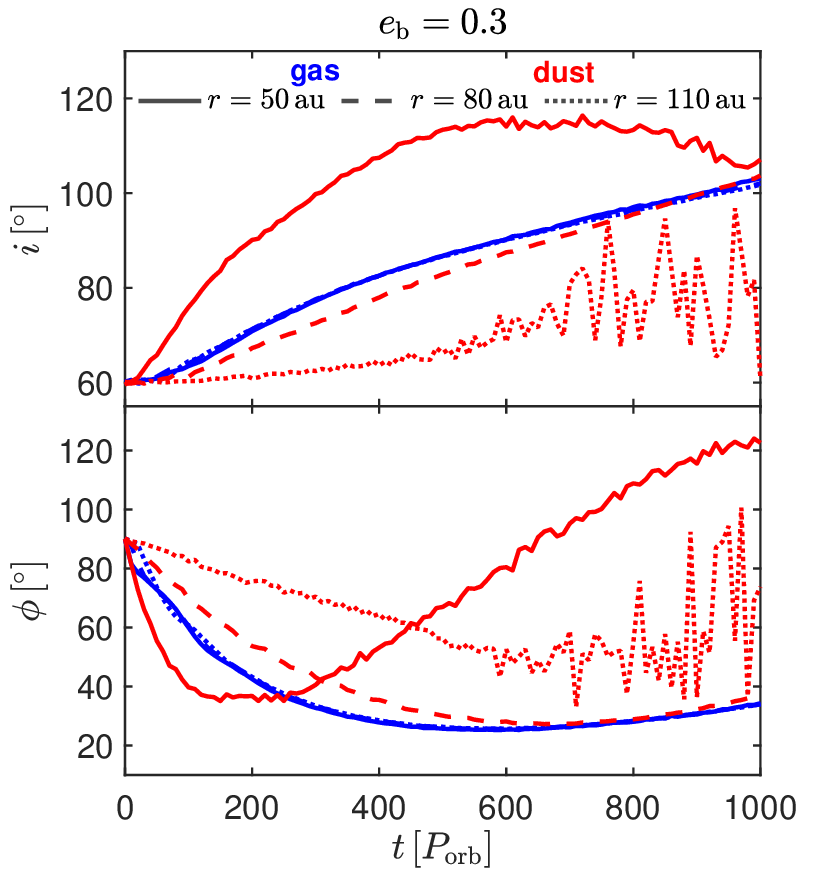}
\includegraphics[width=\columnwidth]{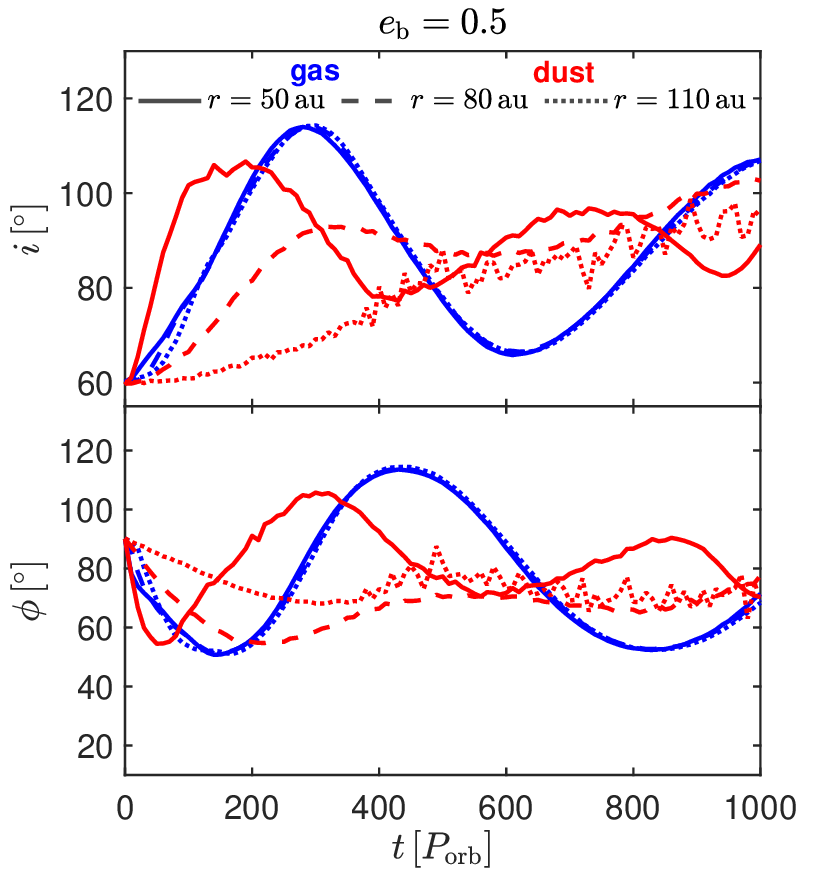}
\centering
\caption{The evolution of the disc tilt $i$ and longitude of the ascending node $\phi$  as a function of time in units of the binary orbital period $P_{\rm orb}$ for binary eccentricities $e_{\rm b} =0.3$ (left panel, run5) and $e_{\rm b} =0.5$ (right panel, run6). The gas is given by the blue curves, and the dust is given by the red curves. We probe the disc at $50\, \rm au$ (solid), $80\, \rm au$ (dashed), and $110\, \rm au$ (dotted).}
\label{fig::disc_params_ecc}
\end{figure*}


Since we have dust rings that are produced by dust traffic jams within a misaligned circumbinary disc at different Stokes number, we can examine the dynamics occurring within the rings. An important parameter in the formation of planetesimals via the steaming instability is the midplane dust-to-gas ratio, $\epsilon$. In numerical simulations, when $\epsilon > 1$, the streaming instability can drive strong clumping \cite[e.g.,][]{Johansen2009,Schafer2017,Nesvorny2019,Li2021}. In each of our SPH simulations we calculate the dust-to-gas ratio at the disc midplane. Figure~\ref{fig::mid_St} shows the azimuthally-averaged disc midplane dust-to-gas ratio as a function of disc radius and time for $\rm St \sim 15$ (top-left panel), $\rm St \sim 30$ (top-right panel), $\rm St \sim 65$ (bottom-left panel), and $\rm St \sim 100$ (bottom-right panel). In each model,  $\epsilon$ is heightened within the dust rings, which may have important implications for the formation of planetesimals via the streaming instability. In Paper II, we conduct analytical calculations of the streaming instability growth rates given the parameters from the SPH simulations. 

To further understand how the initial dust traffic jam evolves in time, we apply a tracing technique for Figs.~\ref{fig::sigma_St} and \ref{fig::mid_St}. Looking back first at Fig.~\ref{fig::sigma_St}, the dust rings drift inward for each different Stokes number model. In each model, we trace the peak density along the initial dust ring. By taking the gradient of the curve, we obtain a drift rate. Figure~\ref{fig::drift_St} shows the drift rate as a function of time for each Stokes number simulation. The disc with the lowest Stokes number, $\rm St \sim 15$, has the highest drift rate, while the disc with the highest Stokes number, $\rm St \sim 100$, has the lowest drift rate. In each model, the drift rate decreases in time, until the dust ring stalls out in the inner region of the circumbinary disc. We detail in Appendix~\ref{app::dusttrace} that it is a material ring that is drifting rather than a feature in the flow that is moving.

Looking back now at Fig.~\ref{fig::mid_St}, we trace the midplane dust-to-gas ratio, $\epsilon$, along the initial dust ring for each model of varying Stokes number. Figure~\ref{fig::mid_time_st} shows the resulting time evolution of $\epsilon$ at the location of the initial dust ring for each model. The black curve ($\rm St\sim 65$) is the same as in Fig.~\ref{fig::mid_time}, but truncated at $1000\, \rm P_{orb}$ for comparison. As the Stokes number increases, the rate-of-change of $\epsilon$ decreases. Within $1000\, \rm P_{orb}$ the model with $\rm St \sim 15$ reaches above unity, while all other models are less than unity. However, if we extend the simulations, eventually $\epsilon$ may reach above unity for all Stokes numbers (refer back to Fig.~\ref{fig::mid_time}), which bears important implications for the formation of planetesimals via the streaming instability.


\begin{figure*} 
\centering
\includegraphics[width=\columnwidth]{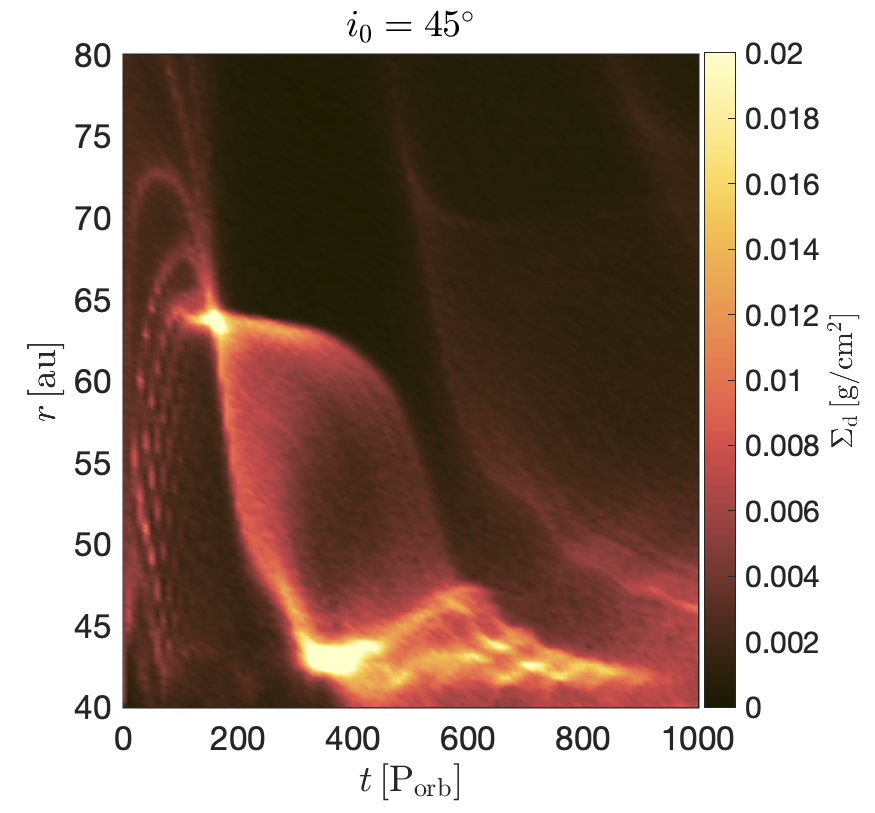}
\includegraphics[width=\columnwidth]{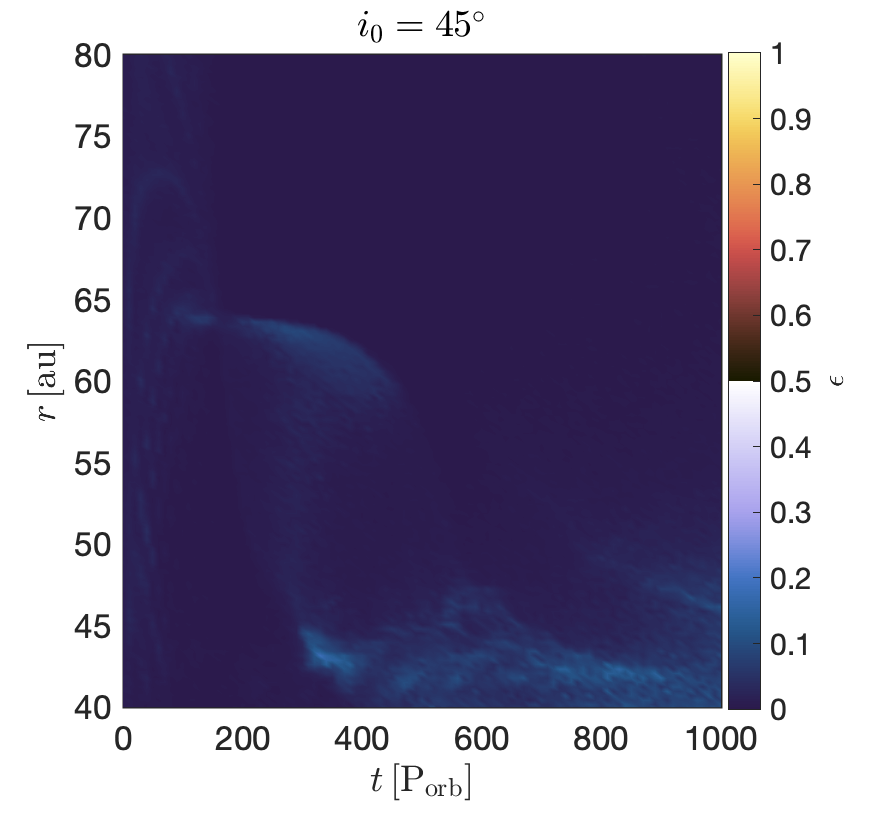}
\centering
\caption{{\it Left panel}: the circumbinary disc dust surface density, $\Sigma_{\rm d}$, as a function disc radius, $r$, and time in binary orbital periods, $\rm P_{orb}$, for a initial disc tilt $i_0 = 45^\circ$ (run7). {\it Right panel}: the circumbinary disc midplane dust-to-gas ratio, $\rho_{\rm d}/\rho_{\rm g}$. }
\label{fig::sigma_inc}
\end{figure*}

\begin{figure} 
\centering
\includegraphics[width=\columnwidth]{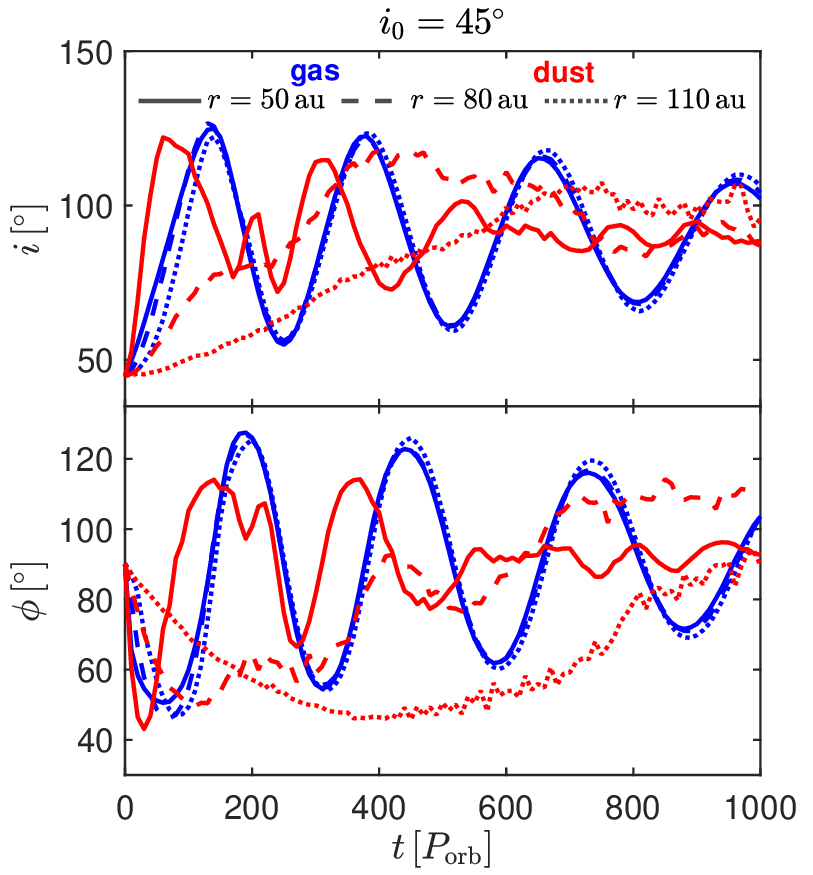}
\centering
\caption{The evolution of the disc tilt $i$ and longitude of the ascending node $\phi$  as a function of time in units of the binary orbital period $P_{\rm orb}$ for initial disc tilt  $i_{0} =45^\circ$ (run7). The gas is given by the blue curves, and the dust is given by the red curves. We probe the disc at $50\, \rm au$ (solid), $80\, \rm au$ (dashed), and $110\, \rm au$ (dotted).}
\label{fig::time_St65_i45}
\end{figure}

\subsection{Varying binary eccentricity}
 We investigate how the binary eccentricity affects the formation of the dust traffic jams. Our standard simulation described in the previous section has a binary eccentricity $e_{\rm b} = 0.8$ with $\rm St \sim 65$. We consider two other binary eccentricities, $e_{\rm b} = 0.3$ (run5) and $0.5$ (run6), both with $\rm St\sim 65$. The left panel in Fig.~\ref{fig::sigma_ecc} shows the azimuthally-averaged disc surface density as a function of time for $e_{\rm b} = 0.5$.  A single dust traffic jam is produced at $t\sim 450\, \rm P_{orb}$. The dust ring drifts inward on a faster timescale compared to our standard simulation. There is no evidence of a secondary dust traffic jam being produced within $1000\, \rm P_{orb}$. The right panel in Fig.~\ref{fig::sigma_ecc}  shows the azimuthally-averaged disc midplane dust-to-gas ratio, $\epsilon$,  as a function of time for $e_{\rm b} = 0.5$. As the dust ring drifts inward, the value of $\epsilon$ increases.

The polar alignment timescale is dependent on the binary eccentricity \citep{Smallwood2020}. For larger binary eccentricities, alignment will occur on a faster timescale, which means a higher frequency of tilt oscillations in a given amount of time. Figure~\ref{fig::disc_params_ecc} shows the disc tilt and longitude of the ascending node as a function of time  for $e_{\rm b} = 0.3$ (left panel) and $e_{\rm b} = 0.5$ (right panel). The gas is given by the blue lines, while the dust is given by the red lines. We probe the disc at three different radial distances, $r = 50$, $80$, $110\, \rm au$.  For $e_{\rm b} = 0.3$, the gas and dust have not undergone one oscillation in tilt or longitude of the ascending node within $1000\, \rm P_{orb}$ due to the weaker binary torque which leads to a weaker precession frequency. This is why the $e_{\rm b} = 0.3$ simulation does not produce any dust traffic jam, $1000\, \rm P_{orb}$ is not a long enough time to trigger a large enough difference in the gas and dust precession in order to produce a dust traffic jam. If the simulation is ran for a long enough timescale, dust traffic jams will form. For $e_{\rm b}=0.5$, there are almost two full oscillations in tilt and longitude of ascending node. The difference in the gas and dust precession produces a dust traffic jam shown previously in Fig.~\ref{fig::sigma_ecc}.

\subsection{Varying disc tilt}
 We investigate how the initial circumbinary disc misalignment affects the formation of the dust traffic jams. Our standard simulation has an initial disc tilt of $i_0 = 60^\circ$ with $\rm St \sim 65$. We consider another initial disc tilt, $i_0 = 45^\circ$ (run7), with $\rm St\sim 65$. The left panel in Fig.~\ref{fig::sigma_inc} shows the azimuthally-averaged disc surface density as a function of radius and time for $i_0 = 45^\circ$. In this case, an initial dust ring is formed at $t\sim 200\, \rm P_{orb}$ at a radius of $\sim 60\, \rm au$ (roughly the same location as the previous simulations). The initial dust ring drifts quickly inward and reaches near the inner disc edge within a few hundred binary orbits. A secondary dust ring is then produced at $t\sim 550\, \rm P_{orb}$. This secondary dust ring forms further out in the disc than the initial dust ring at about $\sim 70\,\rm au$. The secondary dust ring also drifts inward on a fast timescale. Lastly, a third dust traffic jam is produced at a radius $\sim 80\, \rm au$ at a time $t\sim 900\, \rm P_{orb}$. The third dust ring also begins to drift inward. The right panel in Fig.~\ref{fig::sigma_inc} shows the azimuthally-averaged disc midplane dust-to-gas  ratio, $\epsilon$, as a function of radius and time. In this model, $\epsilon$  slightly increases when the dust rings drift to the inner disc edge. Figure~\ref{fig::time_St65_i45} shows the disc tilt and longitude of the ascending node as a function of time. The gas is given by the blue lines, while the dust is given by the red lines. We probe the disc a three different radial distances, $r = 50$, $80$, $110\, \rm au$. The gas and dust are still undergoing polar alignment and are not fully aligned within $1000\, \rm P_{orb}$. The dust undergoes tilt oscillations like the gas, but the alignment timescale for the dust is much longer for dust located in the outer parts of the disc. There is clear difference in the precession between the gas and dust seen by the evolution of the longitude of the ascending node.

\begin{figure*}
\centering
\includegraphics[width=0.50\columnwidth]{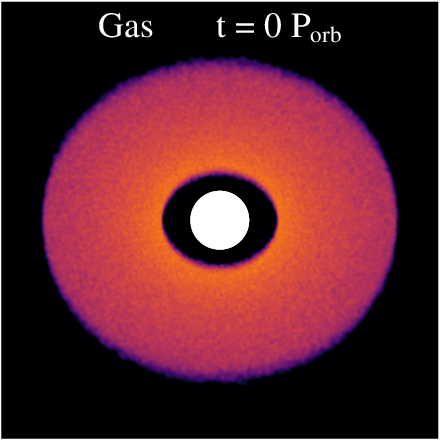}\centering
\includegraphics[width=0.50\columnwidth]{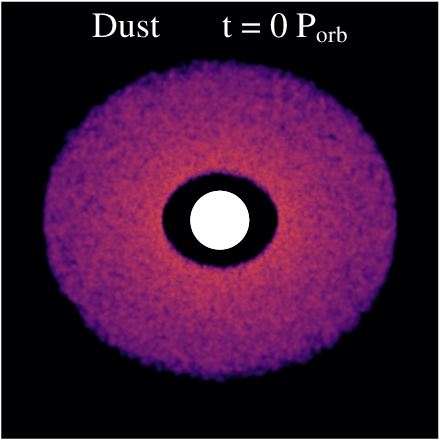}\centering
\includegraphics[width=0.50\columnwidth]{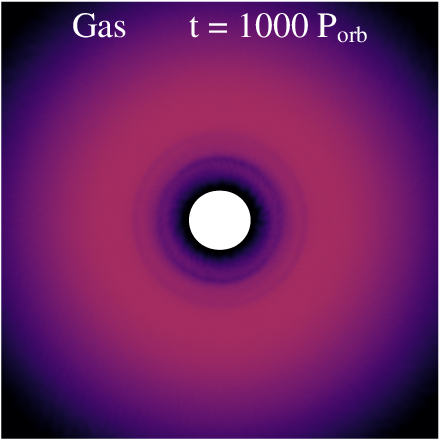}\centering
\includegraphics[width=0.50\columnwidth]{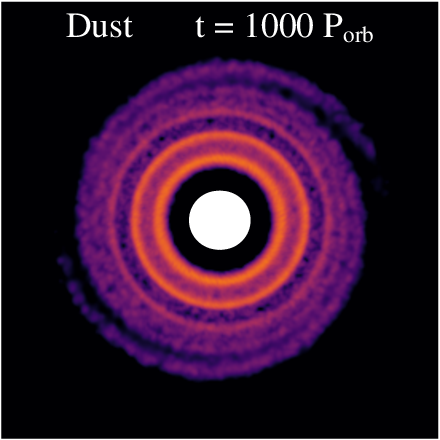}
\includegraphics[width=0.50\columnwidth]{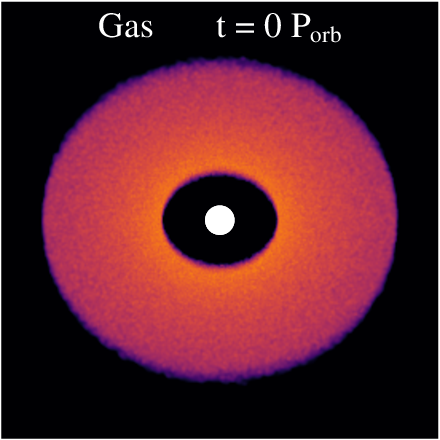}\centering
\includegraphics[width=0.50\columnwidth]{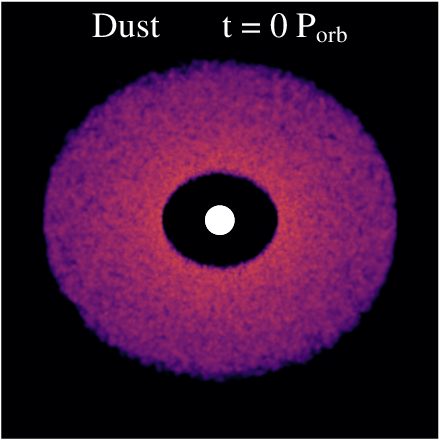}\centering
\includegraphics[width=0.50\columnwidth]{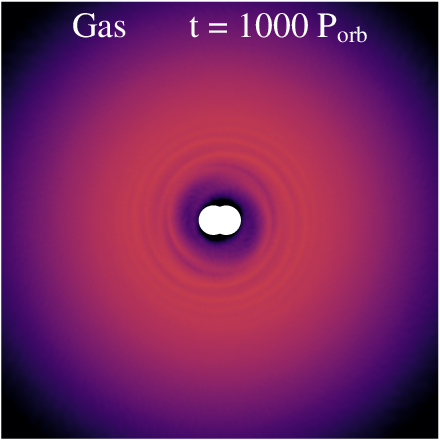}\centering
\includegraphics[width=0.50\columnwidth]{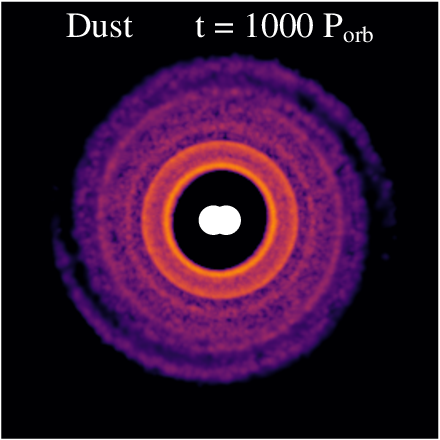}
\includegraphics[width=0.50\columnwidth]{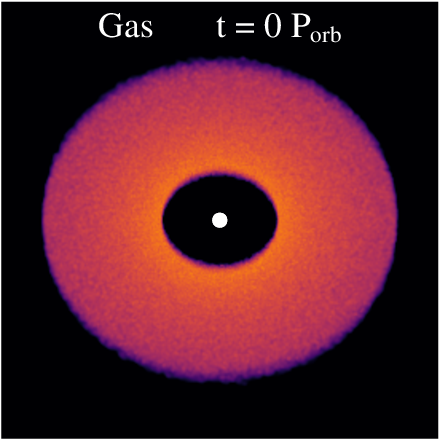}\centering
\includegraphics[width=0.50\columnwidth]{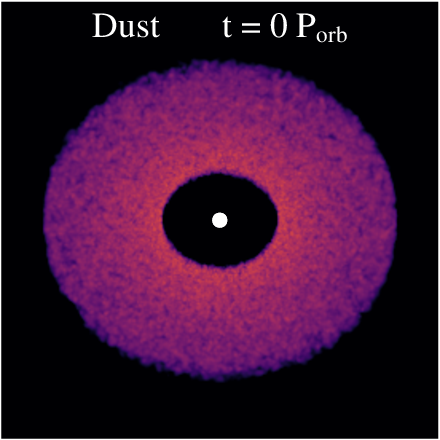}\centering
\includegraphics[width=0.50\columnwidth]{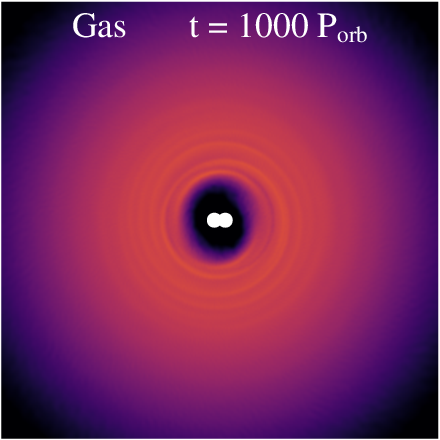}\centering
\includegraphics[width=0.50\columnwidth]{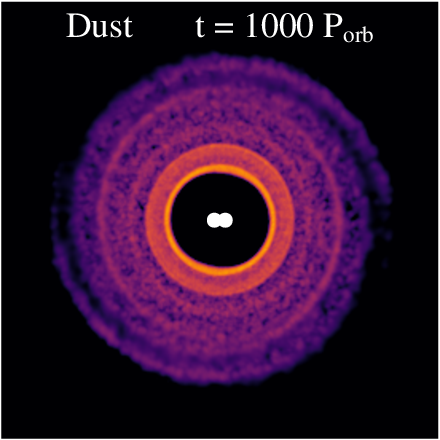}
\includegraphics[width=2\columnwidth]{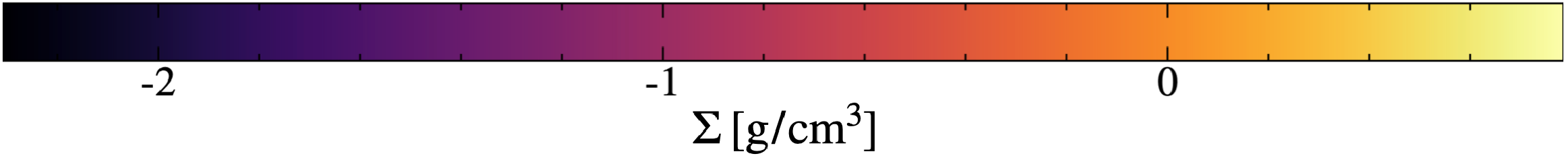}\centering
\caption{The structure and evolution of a highly inclined dusty circumbinary disc with $r_{\rm acc} = 20\, \rm au$ (top row), $10\, \rm au$ (middle row), and $5\, \rm au$ (bottom row). Columns one and two show the initial disc structure for the gas and dust, respectively. Columns three and four show the disc structure at $t = 850\, \rm P_{orb}$. All images are viewed in the $y$-$z$ plane.  The larger accretion radius simulation truncates the inner edge of the disc more than the smaller accretion radius simulations. Furthermore, the surface density of the gas is more depleted in the higher accretion radius simulation.}
\label{fig::splash_resolution}
\end{figure*}

\subsection{Resolution study of the sink accretion radius}
\label{sec::resolution}


In SPH simulations, the sink accretion radius is a parameter used to define the region around a massive object (e.g., a star or planet) where material is gravitationally accreted and added to the object's mass and angular momentum \citep{Bate1995}. This region is often represented by a sink particle, which can gravitationally interact with the surrounding SPH particles in the simulation. Here, we compare the disc structure with three different accretion radii, $20\, \rm au$, $10\, \rm au$, and $5\, \rm au$. 

\begin{figure} 
\centering
\includegraphics[width=\columnwidth]{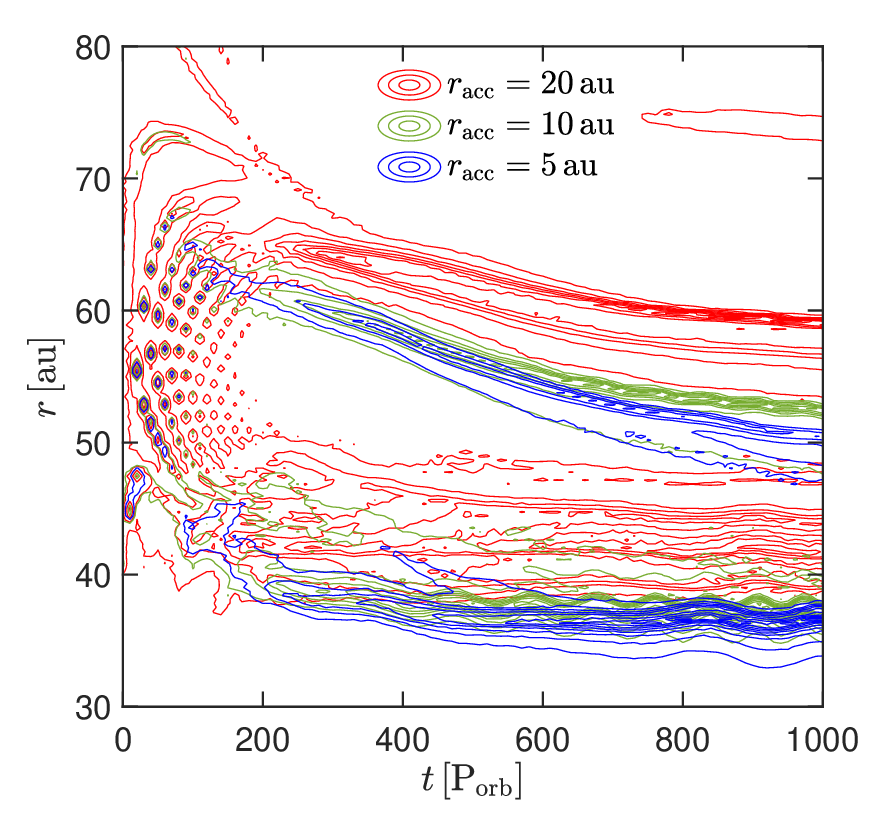}
\centering
\caption{A comparison of the dust ring structure for $r_{\rm acc} = 20\, \rm au$ (red), $r_{\rm acc} = 10\, \rm au$ (green), and $5\, \rm au$ (blue). The contour shows the surface density of the dust. We show the first 10 contours. The location of the dust rings for $r_{\rm acc} = 10\, \rm au$  and $r_{\rm acc} = 5\, \rm au$ are shifted inward due to the binary truncating the disc less than the larger accretion radius simulation. However, the structure and evolution of the resulting dust rings are qualitatively similar.}
\label{fig::resolution}
\end{figure}

Figure~\ref{fig::splash_resolution} shows a comparison of the disc structure of an initially highly inclined gaseous and dusty circumbinary disc with the different sink accretion radii, $r_{\rm acc} = 20\, \rm au$ (top row), $10\, \rm au$ (middle row), and $5\, \rm au$ (bottom row). At $t=1000\, \rm P_{orb}$, the disc aligns to a polar configuration. The larger accretion radius simulation will accrete more of the surrounding disc material, leading to a higher mass accretion rate onto the sink particle, and therefore, deplete the gas disc mass more quickly. Also, the larger accretion radius will more heavily truncate the inner edge of the disc. In a polar-aligned or misaligned disc, the inner edge can lie closer to the binary due to the tidal torque produced by the binary is much weaker \cite[e.g.,][]{Lubow2015,Miranda2015,Nixon2015,Lubow2018,Franchini2019b}, which can be seen in the smaller accretion radii simulations. The inner disc structure is similar with $r_{\rm acc} = 10\, \rm au$ and $5\, \rm au$. However, in each simulation, due to the differential precession of the gas and dust, dust rings are produced in the disc which are qualitatively similar to one another. Figure~\ref{fig::resolution} shows the contours of the dust surface density for $r_{\rm acc} = 20\, \rm au$ (red), $10\, \rm au$ (green), and $5\, \rm au$ (blue). For $r_{\rm acc} = 10\, \rm au$ and $5\, \rm au$, since the inner edge of the disc can lie closer to the binary, the location of the dust ring and the pileup near the inner edge is shifted inward compared to the $r_{\rm acc} = 20\, \rm au$ simulation. 

\begin{figure}
\centering
\includegraphics[width=1\columnwidth]{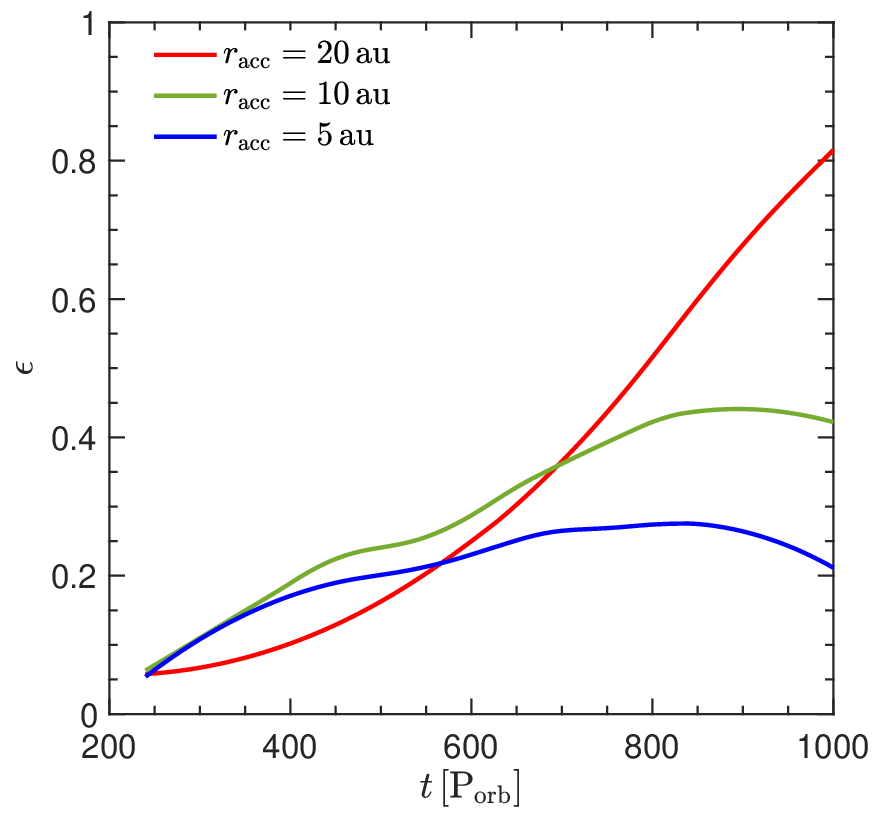}
\caption{ The midplane dust-to-gas ratio, $\epsilon$, as a function of time in binary orbital periods, $\rm P_{\rm orb}$, for the initial dust ring for different sink accretion radii: $r_{\rm acc} = 20\, \rm au$ (red), $r_{\rm acc} = 10\, \rm au$ (green), and $r_{\rm acc} = 5\, \rm au$ (blue).}
\label{fig::midplane_res}
\end{figure}


In our primary suite of simulations with a sink accretion radius $r_{\rm acc} = 20\, \rm au$, the midplane dust-to-gas ratio can surpass unity. From the right panel of Fig.~\ref{fig::disc_params_St65}, the surface density of the initial dust ring does not vary at $1000\, \rm P_{orb}$ to  $2000\, \rm P_{orb}$, but the gas surface density does vary due to accretion of material. This accretion is related to the sink accretion radius. The smaller sink accretion radius, the lower the accretion rate and the more slowly the gas surface density evolves in time. Given these stipulations, the midplane dust-to-gas ratio calculated in the simulations is then dependent on the sink accretion radius. In Fig.~\ref{fig::midplane_res}, we show how the midplane dust-to-gas ratio of the initial dust ring evolves based on the sink accretion radius used in the simulations. The evolution of $\epsilon$ at early times is different between $r_{\rm acc} = 20\, \rm au$ versus the smaller accretion radii due to the difference in the inner disc structure (refer back to Fig.~\ref{fig::splash_resolution}).  For the lower sink accretion radii, $\epsilon$ does not reach above unity within $1000\, \rm P_{orb}$, and begins to decrease. We expect the secondary dust ring will eventually grow and starve the first dust ring \cite[e.g.,][]{Aly2021}.  However, $\epsilon$ is still heightened within the dust rings for the smaller sink accretion radii. In order to observe these dust rings, $\epsilon$ should preferably be below unity. If $\epsilon > 1$ dust growth will quickly occur and the formation of planetesimals will render the dust ring unobservable with the current observational capabilities. Regardless of whether or not $\epsilon$ reaches unity, the formation of dust rings within circumbinary discs may have impact on future circumbinary disc observations and planet formation.

\section{1D calculations}
\label{sec::1D_cal}


\cite{Aly2021} derived the evolution equation for the dust angular momentum in the general case where the gas and dust components are allowed to be inclined with respect to each other (which occurs naturally as a consequence of the varying precession rates between gas and dust), and hence taking into account the drag torque on the evolution. Evolving both gas and dust in the 1D ring code \textsc{RiCo}, they were able to reproduce the dust traffic jams in their SPH simulations for the case of discs around circular binaries (i.e; co-planar alignment). This allowed a straightforward interpretation of the formation of dust traffic jams as a consequence of the drag torque exerted by the gas on the dust which redistributes the dust angular momentum in a way that minimises the radial drift in two locations, inner and outer of the co-precession radius, where the dust traffic jams form.

Here we extend this treatment to the polar alignment case to test whether this simple interpretation is still valid in the eccentric binary case. As in \cite{Aly2021}, the evolution equation for the dust angular momentum is
\begin{equation}
\frac{\partial \bm{L}_\mathrm{d}}{\partial t}  = \bm{T}_\mathrm{d}
-\frac{2}{R} \frac{\partial }{\partial R}\left[ \frac{ R(\bm{T}_\mathrm{d} \cdot \bm{l}_\mathrm{d} )}{\Sigma_\mathrm{d} R \Omega}\bm{L}_\mathrm{d}\right] + \bm{\Omega}_\mathrm{p}\times \bm{L}_\mathrm{d}
\label{equ:angmom_dust}
\end{equation}
Where $\bm{T}_\mathrm{d}$ is the torque density  felt by a dust ring due to drag forces from a gas ring
\begin{equation}
\bm{T}_\mathrm{d}= \frac{k_\mathrm{s}H_\mathrm{g} (\epsilon\bm{L}_\mathrm{g}-\bm{L}_\mathrm{d})}{m_\mathrm{s}(R\upi \sin\theta + H_\mathrm{g})}
\label{equ:drag_torque}
\end{equation}
where $k_\mathrm{s}$ is a drag coefficient, $m_\mathrm{s}$ is the dust particle mass, $\theta$ is the inclination between the gas and dust rings, $H_\mathrm{g}$ is the thickness on the gas ring, $\epsilon=\Sigma_\mathrm{d}/\Sigma_\mathrm{g}$ is the surface density dust-to-gas ratio, 
 and $\bm{L}_\mathrm{g}$ and $\bm{L}_\mathrm{d}$ are the angular momentum densities for gas and dust, respectively. The precession frequency $\bm{\Omega}_\mathrm{p}$ around an eccentric binary is computed from
\begin{equation}
   \bm{\Omega}_\mathrm{p}{(R)}=\frac{3}{4} \frac{\sqrt{G M} \eta a^{2}}{R^{7 / 2}}[5e_\mathrm{b}^2(\hat{\bm{l}}\cdot \hat{\bm{e}})\hat{\bm{e}}-(1-e_\mathrm{b}^2)(\hat{\bm{l}}\cdot \hat{\bm{h}})\hat{\bm{h}}]
    \label{equ:precess_ecc}
\end{equation}
where $\bm{e}$ is the binary eccentricity vector and $\bm{h}$ is the binary specific angular momentum vector. We evolve the gas according to the linearised equations appropriate for the bending wave regime \citep{papaloizou1995,Lubow2000}
\begin{equation}
    \Sigma_\mathrm{g} R^2 \Omega_\mathrm{g}  \frac{\partial \bm{l}_\mathrm{g} }{\partial t} =
\frac{1}{R}\frac{\partial \bm{G}}{\partial R}+\bm{\Omega}_\mathrm{p}\times \bm{L}_\mathrm{g}
\label{equ:gas_1}
\end{equation}
and
\begin{equation}
    \frac{\partial \bm{G}}{\partial t}+ \frac{\Omega_\mathrm{g}^2-\kappa^2}{2\Omega_\mathrm{g}} \, \bm{l}_\mathrm{g}\times \bm{G} + \alpha \Omega_\mathrm{g} \bm{G}=\frac{\Sigma_\mathrm{g} H^2 R^3\Omega_\mathrm{g}^3}{4}\frac{\partial \bm{l}_\mathrm{g}}{\partial R},
\label{equ:gas_2}
\end{equation}
where $\bm{G}$ is the disc internal torque, and $\kappa$ is the epicyclic frequency. Note that the gas surface density is kept fixed in these simulations and that the effects of back-reaction of the dust onto the gas are not included.

\begin{figure} 
\centering
\includegraphics[width=\columnwidth]{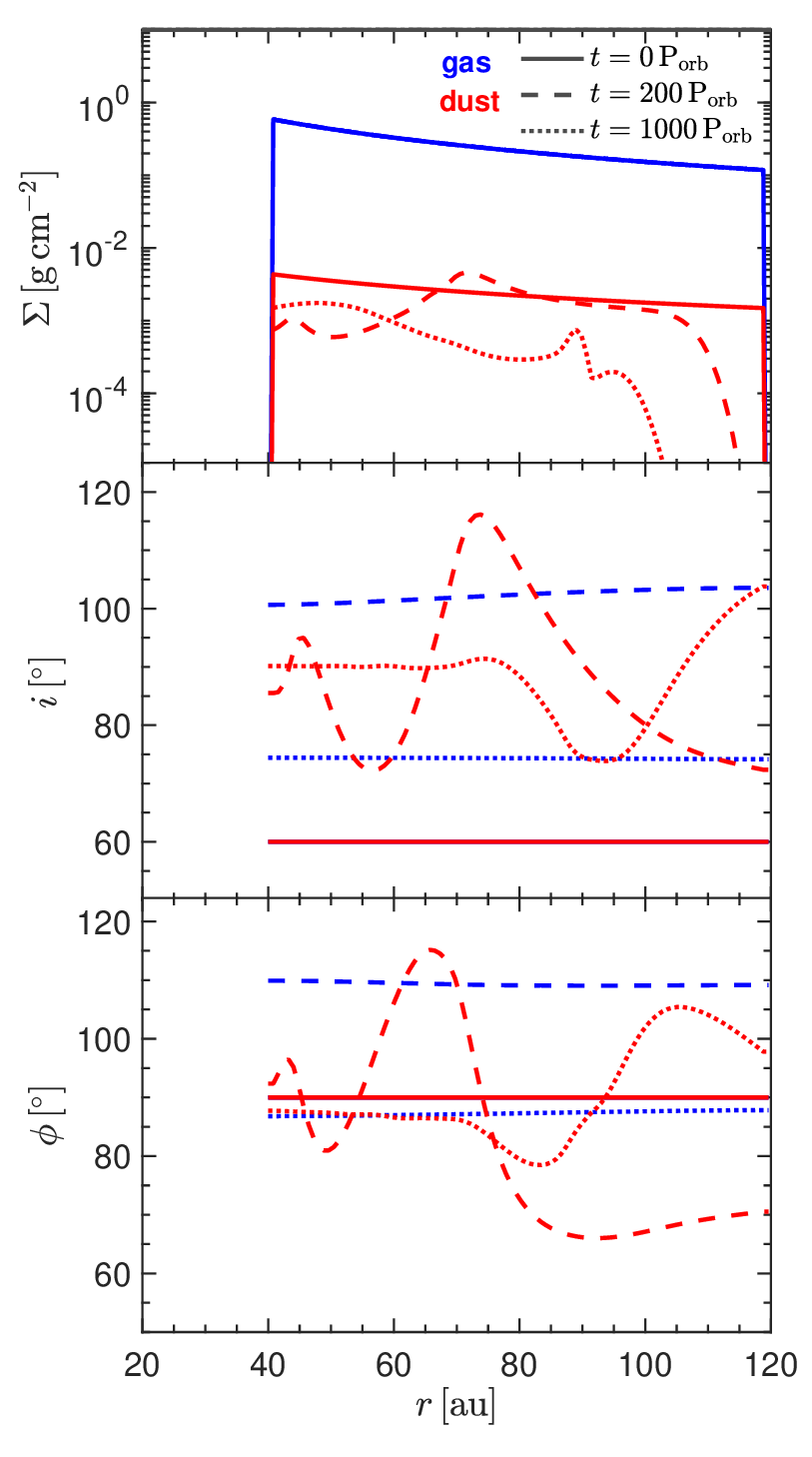}
\centering
\caption{The surface density $\Sigma$ (top panel), tilt $i$ (middle panel), and longitude of ascending node $\phi$ (bottom panel) as a function of disc radius $r$ obtained from the 1D calculations for the standard disc and binary parameters (run4) at $t = 0,\, 200,\, \mathrm{and} \, 1000 \rm P_{orb}$ }
\label{fig::profiles_1d}
\end{figure}

We evolve the above equations using the 1D ring code \textsc{RiCo} \citep{Aly2021} which employs a finite difference, first order, time-explicit scheme and uses a zero-torque boundary condition. The disc is discretised into 100 linearly spaced grid cells. We setup a disc using the same disc and binary parameters of our standard SPH run with $\rm St \sim 65$ (run4). In Figure~\ref{fig::profiles_1d} we show the dust and gas surface density, tilt, and longitude of the ascending node profiles resulting from the 1D calculations at the same evolutionary stages shown in Figure~\ref{fig::disc_params_St65}. Comparing both figures we see that the 1D reproduces both traffic jams in the dust surface density profile (top panels) at similar locations compared to the SPH results. Moreover, the oscillatory evolution, as well as the eventual alignments, of the tilt and twist angles shown in the SPH results of Figure~\ref{fig::disc_params_St65} is well reproduced by the 1D treatment as shown in the middle and bottom panels in Figure~\ref{fig::profiles_1d}.

\section{Discussion}
\label{sec::discussion}

Protoplanetary discs are composed of gas and dust, and their dynamics are influenced by turbulence and various instabilities \citep{Youdin2002}. Dust grains experience gas drag, leading to inward drift due to the differential rotation of the disc \citep{Nakagawa1986}. This process poses challenges to dust growth beyond certain sizes before they are lost to the central star \citep{Birnstiel2012}. Turbulence in protoplanetary disks plays a dual role in the planetesimal formation process \cite[e.g.,][]{Xu2022}. Turbulence can facilitate the diffusion of dust grains, leading to their homogenization throughout the disc. However, turbulence can also create regions of enhanced density, known as dust traps or rings \citep{Johansen2006}. These dust traps arise from the combined effect of radial pressure gradients and turbulent vortices that concentrate dust particles. However, in this work, we showed that dust is not trapped but rather piled-up within a misaligned circumbinary disc. These dust pile-ups or traffic jams will occur naturally within a misaligned circumbinary disc as it undergoes alignment. 

Distinguishing between dust traps and dust traffic jams involves understanding their distinct physical characteristics. In dust traps, dust grains experience reduced inward drift as they are confined within regions of higher density \citep{Zhu2012}. On the other hand, dust is not trapped in a dust traffic jam; instead, the dust grains migrate radially inwards but at a slower pace, resulting in a pile-up of dust density. Notably, dust traps occur at the same pressure maxima (same radii) for all dust species of different sizes, while dust traffic jams manifest at different radii for various dust sizes (refer to Fig.~\ref{fig::sigma_St}, with converging differences at large St). This distinction presents a potential observational avenue to differentiate between dust traps and dust traffic jams, a prospect we aim to explore in future work. Moreover, there are implications for local dust evolution, particularly through mechanisms like streaming instabilities. In a dust trap, radial drift is strictly zero due to the absence of a pressure gradient. In contrast, in a dust traffic jam, there is still radial drift (streaming) of dust through gas, a factor of significance for streaming instabilities.

Dust traffic jams in polar-aligning discs exhibit distinct evolutionary patterns compared to those in coplanar-aligning discs.  A polar-aligning disc precesses at a faster rate compared to a coplanar-aligning disc, leading to quicker formation of dust traffic jams. For example, as demonstrated by \cite{Aly2021}, traffic jams in coplanar-aligning discs typically emerge after $1000, \rm P_{orb}$, whereas in our study, these jams manifest on a faster timescale, within approximately $\sim 200, \rm P_{orb}$. The more extended discs in \cite{Aly2021} take longer for the bending waves to propgate and reach a rigidly precessing regime, compared to the simulations conducted in this work. The dust traffic jams only occur after the gas starts precessing rigidly \citep{Longarini2021,Aly2021}. Furthermore, while traffic jams in polar-aligning discs display an inward drift, those in coplanar-aligning discs remain relatively stationary. This inward drift in polar-aligning discs can have notable implications for planet formation, potentially triggering planetary formation at various radial locations and inclinations. Importantly, the longevity of traffic jams in polar-aligning discs persists beyond the alignment phase. This aspect sets our findings apart from \citep{Aly2021}, as their simulations did not complete the alignment process in the presence of gas or dust. Consequently, our results shed light on the impact of dust traffic jams in polar-aligning discs, emphasizing their relevance in planetesimal growth.

Streaming instabilities, driven by the differential motion between gas and dust, have been identified as key processes responsible for the rapid growth of particles in dust rings \citep{Youdin2005}. These instabilities lead to the formation of dense filaments, further aiding in the concentration and aggregation of dust particles. This concentration enables particles to collide and coagulate more frequently, overcoming typical growth barriers \citep{Brauer2008}. Consequently, larger and more solid aggregates can form within these dust-trapping rings, serving as seeds for the eventual growth of planetesimals. Such a mechanism could explain the presence of large planetesimals in some systems that formed early in the disc's evolution. Observational techniques, such as high-resolution imaging and spectroscopy, have enabled the detection of dust rings in protoplanetary discs \citep{vanderMarel2013}. These observations support the theoretical predictions and emphasize the importance of enhanced dust concentrations in the planetesimal formation process.

In recent times, the Atacama Large Millimeter/submillimeter Array (ALMA) has provided insights into a wide array of structures found in nearby protoplanetary discs. Among these structures, a notable feature is the presence of axisymmetric rings observed in the continuum emission, associated with dust grains \citep{ALMA2015,Andrews2018}. These rings may form due to localized enhancements in gas pressure, or "pressure bumps," which potentially alter the direction of radial drift of solid particles \citep{Whipple1972} or formed via dust traffic jams \citep{Longarini2021,Aly2021}.  The formation of these rings might play a crucial role in augmenting the midplane dust-to-gas ratio (a parameter used in planetesimal formation models), thereby activating the streaming instability process and leading to the formation of planetesimals. Early planetesimal formation, occurring before the influence of photoevaporation becomes significant \citep{Carrera2017}, could be particularly dependent on these structures. 

The exploration of dust traffic jams in misaligned circumbinary discs necessitates a combined approach involving 3D simulations and 1D analysis. Our 1D method effectively identifies and characterizes these traffic jams, affirming that the underlying physics involves the redistribution of angular momentum within the disc. This redistribution, driven by drag torque on the dust, minimizes radial drift at the traffic jam location. Notably, our 1D code accurately captures the dynamic nature of the polar disc case, where the traffic jam is non-stationary, as observed in \cite{Aly2021}. The 'inward drift' of the traffic jam is explained by the evolving tilt and twist profiles of the gas and dust, influencing the location of the minimum radial drift. This process also leads to the creation of new traffic jams over time. It is crucial to highlight that our ability to recover these phenomena with the 1D algorithm unequivocally establishes them as traffic jams, not dust traps. The algorithm's constraint on the radial gas density profile precludes the development of pressure bumps and, consequently, any formation of dust traps. However, while the 1D approach successfully predicts the occurrence and radial evolution of traffic jams, accurately modeling the vertical thickness of the disc, particularly in terms of the extent of dust density enhancement, requires 3D SPH simulations. The 1D method tends to underestimate these enhancements. A significant outcome of our study is the revelation of a remarkably high midplane dust fraction, potentially triggering streaming instabilities, a phenomenon best unveiled through precise 3D modeling.

\section{Summary}
\label{sec:conclusion}
We investigated the formation of dust traffic jams in polar-aligning circumbinary discs. We first used SPH simulations of both gas and dust to model an initially highly misaligned circumbinary disc around an eccentric binary. As the circumbinary disc evolves to a polar state, the difference in the precession between the gas and dust produces dust traffic jams \cite[e.g.,][]{Aly2020,Longarini2021,Aly2021} which become dense dust rings. We found the formation of dust rings exists for different Stokes number, binary eccentricity, and initial disc tilt. These dust rings are only formed when the disc is misaligned to the binary orbital plane. When the disc is polar aligned, the engine to produce the dust rings is switched off.  The evolution of dust rings varies between a polar-aligning disc and a coplanar-aligning disc, with three key distinctions: 1) dust traffic jams form on a quicker timescale  due to a faster precession rate, 2) in polar discs, the dust traffic jams drift inward instead of remaining stationary, and 3) they persist beyond the alignment phase.

However, if the dust rings are formed before the disc aligns polar, the rings are still able to survive in a polar state. Once these dust rings are formed, they drift inward. The drift timescale is faster for a lower Stokes number. The majority of the simulations showed a secondary dust ring that is produced at later times. Next, we demonstrated that the 1D model from \cite{Aly2021} produced dust rings in a misaligned circumbinary disc that is undergoing polar alignment. The formation of stable polar dust rings has implications for forming polar planets. From our SPH simulations,  these dust rings will have an increased midplane dust-to-go ratio, which may be a favourable environment for the steaming instability to operate.

\section*{Acknowledgements}
 We thank the anonymous referee for helpful suggestions that improved the quality of the manuscript. JLS thanks Rebecca Martin for helpful discussions. We acknowledge the use of Sarracen \citep{Harris2023} for rendering some of the figures. JLS acknowledges funding from the ASIAA Distinguished Postdoctoral Fellowship. MKL is supported by the National Science and Technology Council
(grants 111-2112-M-001-062-, 112-2112-M-001-064-, 111-2124-M-002-013-, 112-2124-M-002 -003-) and an Academia Sinica Career Development Award (AS-CDA110-M06). HA acknowledges funding from the European Research Council (ERC) under the European Union’s Horizon 2020 research and innovation programme (grant agreement No 101054502). RN acknowledges funding from UKRI/EPSRC through a Stephen Hawking Fellowship (EP/T017287/1). CL acknowledges funding from the European Union’s Horizon 2020 research and innovation programme under the Marie Skłodowska-Curie grant agreement No 823823 (RISE DUSTBUSTERS project). 

\section*{Data Availability}
The data supporting the plots within this article are available on reasonable request to the corresponding author. A public version of the {\sc phantom}, {\sc splash}, and {\sc sarracen} codes are available at \url{https://github.com/danieljprice/phantom}, \url{http://users.monash.edu.au/~dprice/splash/download.html}, and \url{https://sarracen.readthedocs.io/en/latest/index.html}  respectively. RiCo is available upon reasonable request to Rebecca Nealon.


\bibliographystyle{mnras}
\bibliography{ref.bib} 




 \appendix

\section{Extended simulation of the dust ring}
\label{app::extended}

\begin{figure*}
\centering
\includegraphics[width=1\columnwidth]{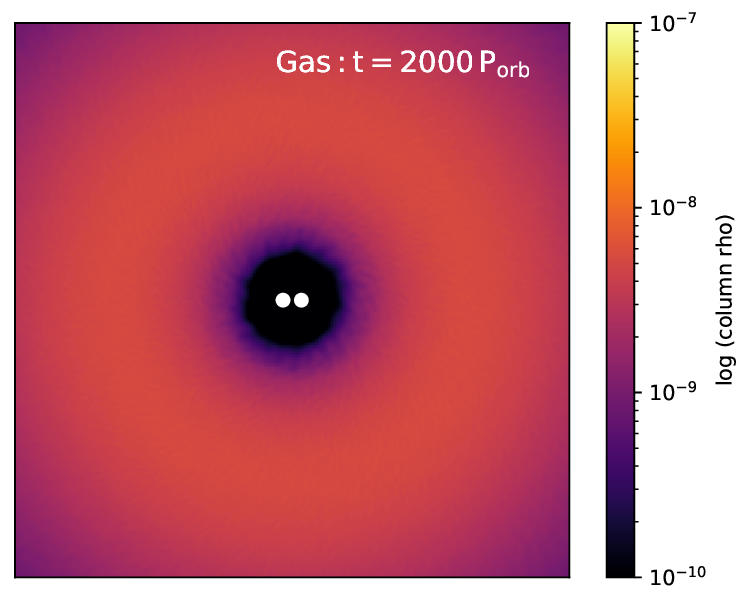}
\includegraphics[width=1\columnwidth]{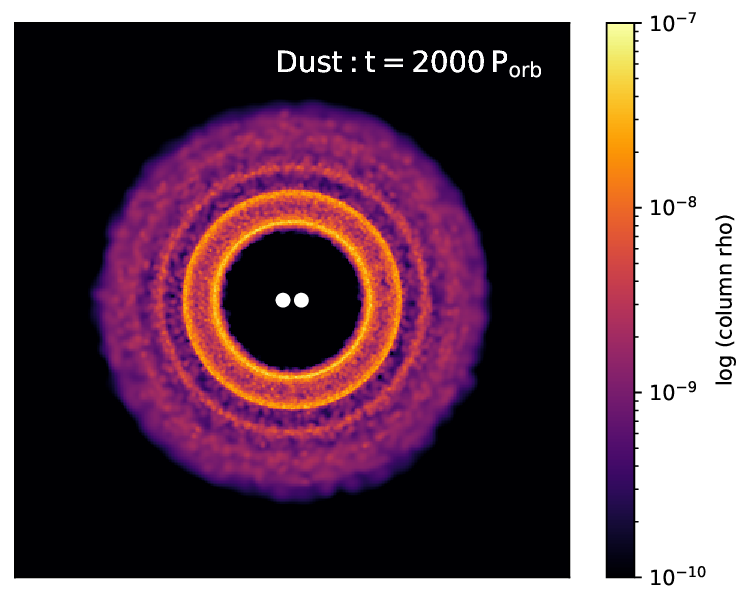}
\caption{The structure of the gas (left panel) and dust (right panel) disc at $t = 2000\, \rm P_{orb}$. The colour bar denotes the surface density. The image is viewed in the $y$--$z$ plane.  }
\label{fig::t2000_image}
\end{figure*}

Figure~\ref{fig::t2000_image} shows the structure of the gas (left panel) and dust (right panel) in the circumbinary disc for your standard simulation (run1) at $t = 2000\, \rm P_{orb}$. There are no substructures in the gas surface density, but there are multiple rings present in the dust.

 \section{Dust traffic jam to dust ring}
 \label{app::dustring}

 \begin{figure*}
\centering
\includegraphics[width=0.50\columnwidth]{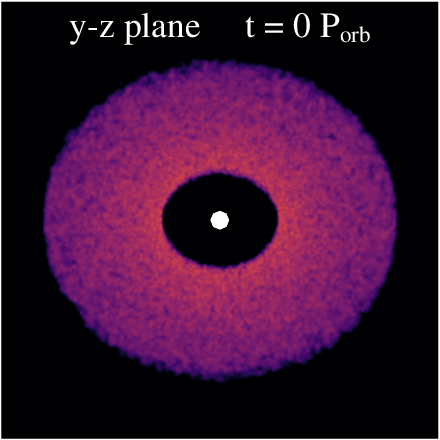}\centering
\includegraphics[width=0.50\columnwidth]{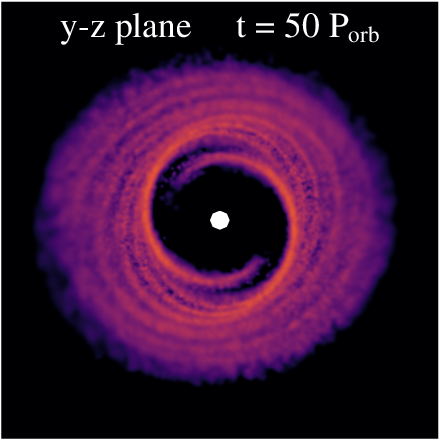}\centering
\includegraphics[width=0.50\columnwidth]{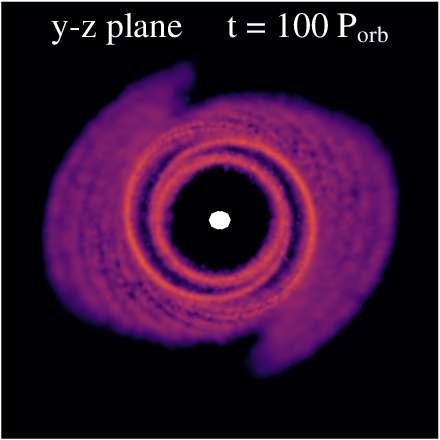}\centering
\includegraphics[width=0.50\columnwidth]{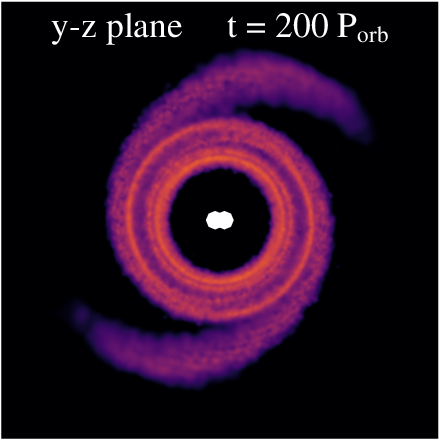}
\includegraphics[width=2\columnwidth]{plots/colorbar.eps}\centering
\caption{Dust surface density snapshots showing the evolution from dust traffic jam to a coherent dust ring. We show the snapshots at times $t = 0\, \rm P_{orb}$, $50\, \rm P_{orb}$, $100\, \rm P_{orb}$, and $200\, \rm P_{orb}$. The disc is viewed in the $y$--$z$ plane. }
\label{fig::splash_dustring}
\end{figure*}

In a misaligned circumbinary disc, the differential precession of the gas and dust create regions of enhanced dust concentration or "dust traps."  When dust particles experience drag forces from the gas in the disc, they lose angular momentum and start to migrate towards the central binary stars. However, when they reach the dust trap regions, their migration stalls, forming a dust traffic jam. As dust particles accumulate in the dust traps, they create regions of increased surface density within the circumbinary disc. This accumulation leads to the formation of localized overdensities or clumps of dust. Over time, the accumulated dust in the traffic jam region increases, forming a more coherent and compact structure known as a dust ring. The dust ring is a ring-shaped or annular structure that encircles the binary stars. We show snapshots of this process of the dust traffic jam becoming a dust ring in Fig.~\ref{fig::splash_dustring}. The dust ring can serve as a favorable environment for the coagulation and collisional growth of dust particles, eventually leading to the formation of planetesimals, the building blocks of planets.

\section{Dust ring location}
\label{app:ring_location}

\begin{figure} 
\centering
\includegraphics[width=\columnwidth]{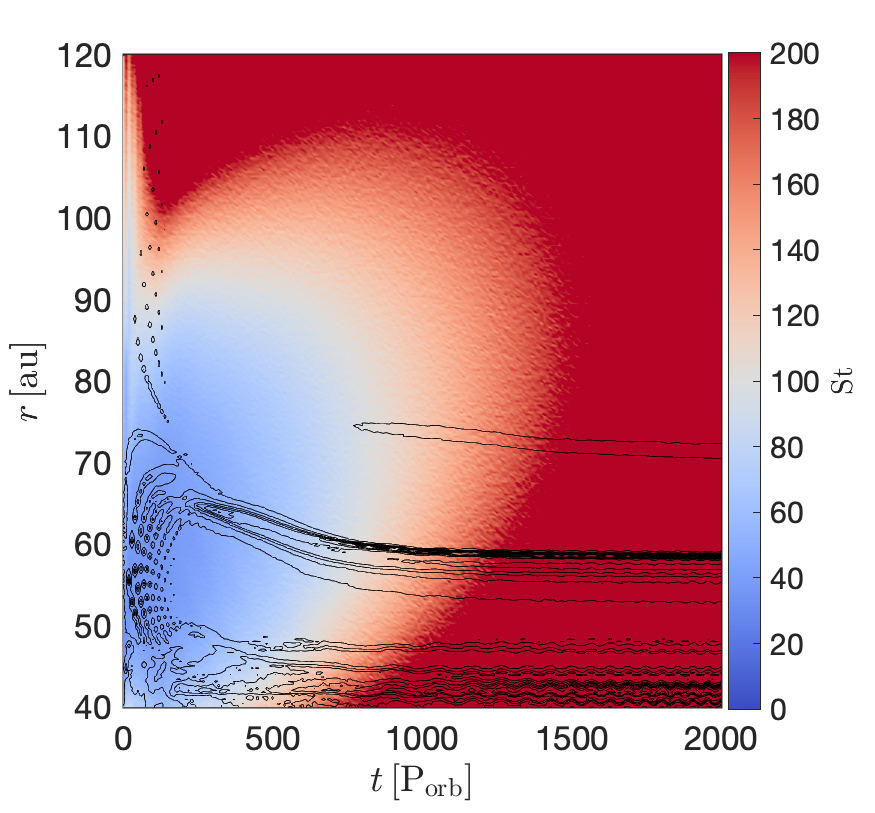}
\centering
\caption{ The azimuthally-averaged Stokes number has a function of radius, $r$, and time in binary orbital periods, $\rm P_{\rm orb}$ for our standard simulation (run1) with an initial Stokes number of $65$. The Stokes number is calculated from Eq.~\ref{eq::st}. The black contour shows the first ten contour levels of the dust surface, which traces the dust rings. The Stokes number of the grains in the disc change overtime due to the evolving surface density of the gas, which halts the inward drift of the dust rings.}
\label{fig::st_65}
\end{figure}

 Figure~\ref{fig::st_65} shows the azimuthally-averaged Stokes number has a function of radius, $r$, and time in binary orbital periods, $P_{\rm orb}$ for our standard simulation (run1) with an initial Stokes number of $65$. The Stokes number is calculated from Eq.~\ref{eq::st}, which depends on the surface density of the gas. The black contour shows the first ten contour levels of the surface density of the dust ring. The Stokes number of the grains in the disc change overtime due to the evolving surface density of the gas. Over time, the Stokes number of the grains in the disc changes due to the evolving surface density of the gas. As material accretes onto the binary, causing a decrease in the gas surface density, the Stokes number increases, given that the dust grains in our simulations neither grow nor fragment. The initial dust ring migrates inward when the Stokes number is relatively low. However, as the Stokes number in the disc increases, the inward drift slows down. Eventually, when the Stokes number reaches sufficiently high values, the dust grains become more decoupled from the gas, effectively halting their inward drift.

\section{Inward drift of the Dust ring}
\label{app::dusttrace}
\begin{figure}
\centering
\includegraphics[width=1\columnwidth]{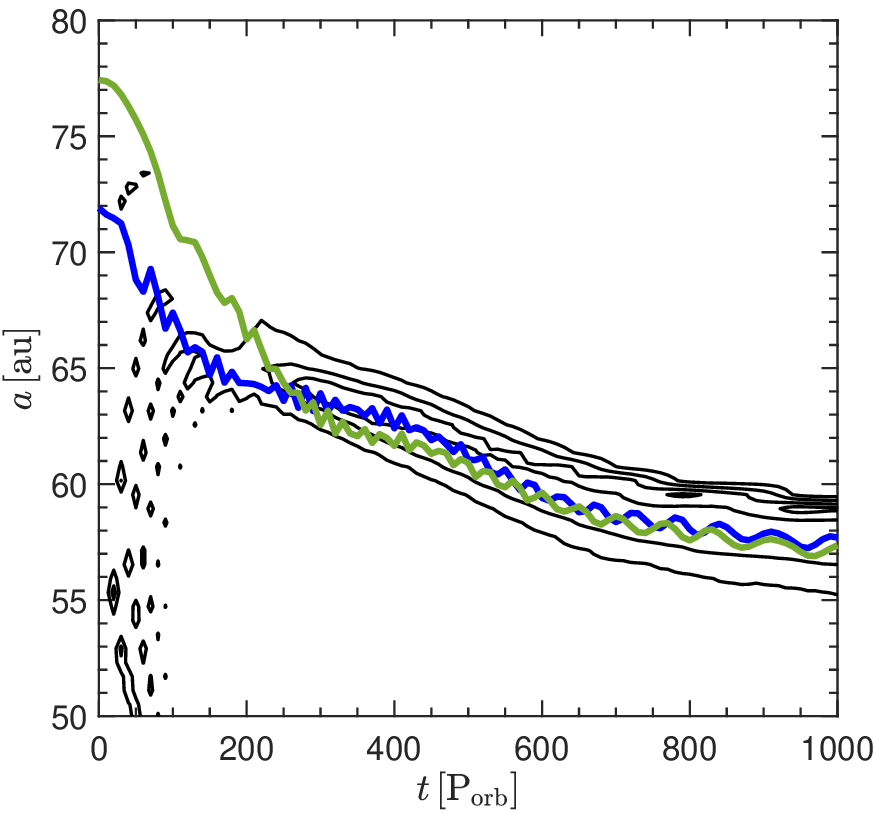}
\caption{The semi-major axis, $a$, of two dust particles as a function of time in binary orbital periods, $\rm P_{orb}$. The black contour shows the surface density of the dust ring. We show five contour levels. The dust particles drift at the same rate as the dust ring, signifying that the dust ring drifts as a material ring rather than a feature of the flow.}
\label{fig::trace_plot}
\end{figure}

We are investigating whether the observed movement of the material ring is a consequent of the entire ring drifting or rather a feature of the flow within the disc. We track the trajectories of two dust particles that eventually end up inside the initial dust ring.  In Fig.~\ref{fig::trace_plot}, we display the semi-major axis of these two dust particles as a function of time, overlaid with the dust surface density contour in black. The two selected dust particles are initially outside of the initial location of the dust ring. Once the dust particles drift into the ring,  they show the same drifting rate as the dust ring, suggesting that the dust ring is drifting as a cohesive material ring rather than being a result of a feature in the flow.

\bsp	
\label{lastpage}
\end{document}